%% file: main.tex
\documentclass[conference]{IEEEtran}

\usepackage{0a_preamble}

\begin{document}

\setlength{\abovedisplayskip}{6pt}
\setlength{\belowdisplayskip}{6pt}

\input{00_macros}

\input{0c_title.tex}

%\IEEEpeerreviewmaketitle
\maketitle

\input{00_abstract.tex}
\input{01_Introduction.tex}
\input{02a_Related_works.tex}
\input{02_Preliminaries.tex}
\input{03_Attack_Objectives.tex}

\input{04_Attack_Design.tex}
\input{05_Victim_Parameter_Sensing.tex}
\input{06_Large_Scale_Evaluation.tex}
\input{07_Physical_Implementation}
\input{07a_Real_World_Case_Studies}
%\input{07b_mmWave_Case_Studies.tex}
\input{08_Discussion_and_future_works.tex}
\input{09_Conclusion.tex}

\input{A0_acknowledgments.tex}

\bibliographystyle{IEEEtran}

%\bibliography{references/references}

\appendix
\input{A1_Appendix_1.tex}
\input{A2_Appendix_2.tex}

\input{A3_Appendix_3.tex}

\end{document}

%% file: 00_macros.tex
%% paper-specific variable/notation
\newcommand\name{MadRadar}
\newcommand\namebf{\textbf{\name}}

%% units
%power
\newcommand{\dBsm}[1]{{#1}\thinspace{dBsm}}
\newcommand{\dBsmSq}[1]{{#1}\thinspace{$\textrm{dBsm}^2$}}
\newcommand{\dB}[1]{{#1}\thinspace{dB}}
\newcommand{\dBm}[1]{{#1}\thinspace{dBm}}
%times
\newcommand{\nsec}[1]{{#1}\thinspace{ns}}
\newcommand{\usec}[1]{{#1}\thinspace{$\upmu$s}}
\newcommand{\msec}[1]{{#1}\thinspace{ms}}
\newcommand{\seconds}[1]{{#1}\thinspace{s}}

%frequencies
\newcommand{\GHz}[1]{{#1}\thinspace{GHz}}
\newcommand{\MHz}[1]{{#1}\thinspace{MHz}}
\newcommand{\kHz}[1]{{#1}\thinspace{kHz}}

%slopes
\newcommand{\MHzPerus}[1]{{#1}\thinspace{MHz/$\upmu$s}}
%distances
\newcommand{\m}[1]{{#1}\thinspace{m}}
\newcommand{\cm}[1]{{#1}\thinspace{cm}}
%velocities
\newcommand{\mPers}[1]{{#1}\thinspace{m/s}}

%constants
\newcommand{\LightSpeed}{\textrm{c}}

%% victim notation
\newcommand{\NumChirpsPerFrame}{N_{\textrm{chirps}}}
\newcommand{\FrameDuration}{T_{\textrm{frame}}}
\newcommand{\ChirpDuration}{T_{\textrm{chirp}}}
\newcommand{\ChirpFreqStart}{f_{c}}
\newcommand{\ChirpSlope}{S}
\newcommand{\ChirpSlopeVic}{S_{\textrm{vic}}}
\newcommand{\ChirpSlopeAtt}{S_{\textrm{atk}}}
\newcommand{\ChirpBW}{B}
\newcommand{\FreqIF}{f_{\textrm{IF}}}
\newcommand{\FreqSamp}{f_{\textrm{samp}}}
\newcommand{\NChirps}{N_{\textrm{Chirps}}}

%radar performance
\newcommand{\RangeRes}{d_{\textrm{res}}}
\newcommand{\RangeMax}{d_{\textrm{max}}}
\newcommand{\RangeMin}{d_{\textrm{min}}}
\newcommand{\VelocityRes}{v_{\textrm{res}}}
\newcommand{\VelocityMax}{v_{\textrm{max}}}
\newcommand{\DopplerShift}{\phi_{\textrm{doppler}}}
\newcommand{\tDelay}{t_{\textrm{d}}}

%attack parameters
\newcommand{\EstimatedChirpSlope}{\widetilde{S}}
\newcommand{\EstimatedChirpDuration}{\widetilde{T}_{\textrm{chirp}}}
\newcommand{\EstimatedFrameDuration}{\widetilde{T}_{\textrm{frame}}}
\newcommand{\AttacktDelay}{t'_{\textrm{d}}}
\newcommand{\AttackDopplerShift}{\phi'_{\textrm{doppler},n}}
\newcommand{\AttackDopplerShiftFN}[1]{\phi'_{\textrm{doppler},{#1}}}
\newcommand{\AttackSlopeFN}{\widetilde{S}'}
\newcommand{\AttackAmplitude}{A_{\textrm{atk}}}
\newcommand{\VelocitySpoof}{v_{\textrm{spoof}}}
\newcommand{\VelocityAttacker}{v_{\textrm{atk}}}
\newcommand{\RangeSpoof}{d_{\textrm{spoof}}}
\newcommand{\RangeAttacker}{d_{\textrm{atk}}}

%attack errors
\newcommand{\RangeError}{R_{\textrm{error}}}
\newcommand{\TimingError}{t_{\textrm{error}}}
\newcommand{\RangeSpread}{R_{\textrm{spread}}}
\newcommand{\IFSpread}{IF_{\textrm{spread}}}

% for the target
\newcommand{\RangeTarget}{d_{\textrm{target}}}
\newcommand{\VelocityTarget}{v_{\textrm{target}}}

%common names
\newcommand{\lidar}{LiDAR}

%project name
\newcommand{\ProjectName}{MAD-RAD}

%%comments
%create flags to show comments/todos
\newif\ifShowComments
\newif\ifShowToDos

%set flags
\ShowCommentstrue
\ShowToDostrue

%tingjun comments
\newcommand\tingjun[1]{%
    \ifShowComments
        \textcolor{red}{\todo{[TC: #1]}} 
    \fi%
}

%david comments
\newcommand\david[1]{
    \ifShowComments\textcolor{blue}{\todo{[DH: #1]}} \fi%
}

%david ToDo comments
\newcommand\davidToDo[1]{
    \ifShowToDos\textcolor{green}{[DH-TODO: #1]}\fi%
}

\newcommand\tingjunToDO[1]{
    \ifShowToDos\textcolor{red}{[TODO: #1]}\fi%
}

%Revisions
\newcommand\Revised[1]{
    \textcolor{blue}{#1}
}

%% file: 0c_title.tex
\title{MadRadar: A Black-Box Physical Layer Attack Framework on mmWave Automotive FMCW Radars}

%create flags to show comments/todos
\newif\ifAnonymize

%set flags
\Anonymizefalse

%add title if Anonymize is false
\ifAnonymize

\else
    \author{\IEEEauthorblockN{David Hunt}
    \IEEEauthorblockA{Duke University\\
    david.hunt@duke.edu}
    \and
    \IEEEauthorblockN{Kristen Angell}
    \IEEEauthorblockA{Duke University\\
    kristen.angell@duke.edu}
    \and
    \IEEEauthorblockN{Zhenzhou Qi}
    \IEEEauthorblockA{Duke University\\
    zhenzhou.qi@duke.edu}
    \and
    \IEEEauthorblockN{Tingjun Chen}
    \IEEEauthorblockA{Duke University\\
    tingjun.chen@duke.edu}
    \and
    \IEEEauthorblockN{Miroslav Pajic}
    \IEEEauthorblockA{Duke University\\
    miroslav.pajic@duke.edu}}

%\author{
%\IEEEauthorblockN{David Hunt, Kristen Angell, Zhenzhou Qi, Tingjun Chen, Miroslav Pajic}
%\IEEEauthorblockA{Department of Electrical and Computer Engineering, Duke University \\
%Email: \{david.hunt, krisen.angell, zhenzhou.qi, tingjun.chen, miroslav.pajic\}@duke.edu}
% \thanks{}
%}

\fi

%% file: 00_abstract.tex
\begin{abstract}
Frequency modulated continuous wave (FMCW) millimeter-wave (mmWave) radars play a critical role in many of the advanced driver assistance systems (ADAS) featured on today's vehicles.  While previous works have demonstrated (only) successful false-positive spoofing attacks against these sensors, all but one assumed that an attacker had the runtime knowledge of the victim radar's configuration. In this work, we introduce {\namebf}, a general black-box radar attack framework for automotive mmWave FMCW radars capable of estimating the victim radar's configuration in real-time, and then executing an attack based on the estimates. We evaluate the impact~of~such attacks maliciously manipulating a victim radar's point cloud, and show the novel ability to effectively `add' (i.e., false positive attacks), `remove' (i.e., false negative attacks), or `move' (i.e., translation attacks) object detections from a victim vehicle's scene. Finally, we experimentally demonstrate the feasibility of our attacks on real-world case studies performed using a real-time physical prototype on a software-defined radio platform.
\end{abstract}

%% file: 01_Introduction.tex
\section{Introduction}
\label{Introduction}

Radio detection and ranging (a.k.a., \emph{radar}) sensors have traditionally been popular in the automotive market due to their reliability in adverse lighting and weather conditions, long detection range, and ability to detect an object’s relative velocity~\cite{keysight_how_2020,benjamin_imaging_2019,gardill_automotive_2019}. While various techniques and waveforms can be used to perform radar ranging, frequency modulated continuous wave (FMCW) radars are the most common due to the relatively simple implementation at low cost~\cite{gardill_automotive_2019}. 

The latest generation of automotive radar in the millimeter-wave (mmWave) frequency bands utilizes greater bandwidths in the frequency range of \GHz{76--77} 
%\todo{76-76? @DH -- fix}
(i.e., long-range sensing) and \GHz{77--81} (i.e., short-to-mid range sensing). The higher frequencies and greater bandwidth enable these sensors to have 20$\times$ better range resolution (down to \cm{4}), 3$\times$ greater velocity resolution, and a smaller overall sensor footprint~\cite{keysight_how_2020,ramasubramanian_moving_2017}. Given their traditional benefits and the additional capabilities presented by the latest generation of mmWave radars, FMCW radars play a critical role in many advanced driver assistance systems (ADAS) including blind spot detection (BSD), auto emergency braking systems (AEBS), lane change assist (LCA), and rear traffic alert (RTA) systems~\cite{keysight_how_2020,gardill_automotive_2019}. Moving forward, autonomous driving companies (e.g., Mobileye) also plan to use radar sensors to provide additional sensing and redundancy in their future autonomous vehicles by creating a ``360$^\circ$ Radar cocoon''~\cite{mobileye_radar_2022}. As radars continue to gain popularity in automotive systems and applications, it is imperative to understand the vulnerabilities of these systems.

While there are a plethora of analyses for camera and {\lidar} (light detection and ranging) vulnerabilities in autonomous vehicles (e.g.,~\cite{sun_towards_2020,abdelfattah_adversarial_2021,hallyburton_security_2022,cao_adversarial_2019,hallyburton2023partialinformation}), automotive radars have only recently started to attract attention in the security community. Existing security research dealing with physical layer (PHY) attacks on FMCW radar systems has solely focused on spoofing attacks inserting false points into a victim radar’s point cloud -- i.e., \emph{false positive} (FP) attacks. No \emph{false negative} (FN) attacks, resulting in a `removal' of an existing object from the victim radar's scene, have been demonstrated. Similarly, no prior work has introduced \emph{translation} attacks that can `move' detections of existing objects in the victim radar's scene. Instead, initial works~\cite{sun_who_2021,miura_low-cost_2019,komissarov_spoofing_2021,chauhan_platform_2014} only demonstrated the ability to insert FP objects at a specific range in a radar's point cloud, and more recently showed the ability to spoof an object's velocity~\cite{sun_who_2021,miura_low-cost_2019}. Moreover,~\cite{sun_who_2021} demonstrated the ability to spoof an object's angle of arrival (AoA). However, existing methods, except the very recent one from~\cite{vennam_mmspoof_nodate}, assumed a \emph{white-box threat model with full knowledge} of the victim radar's parameters, significantly limiting their real-world use. 

Additionally,~\cite{chen_metawave_2023} and~\cite{ranganathan_physical-layer_2012} introduced passive attacks and early detect/late commit (ED/LC) attacks, respectively. While~\cite{chen_metawave_2023} introduced \emph{passive} attacks on FMCW radars using physical patches placed in the environment, these attacks are limited as the attacks cannot dynamically change the spoofing location and each patch must be specifically designed for the specific attack goals, victim radar configuration, and environment. The ED/LC attack~\cite{ranganathan_physical-layer_2012} listens to and then re-transmits a victim's signal to spoof an object's range, but the attack is only designed for chirp spread spectrum-based ranging and thus does not work against FMCW radars.

In this work, we present {\namebf}, \emph{a novel real-time \textbf{black-box} FMCW radar attack framework for successful FP, FN, and translation attacks, where an attacker learns the victim radar's parameters and then successfully launches an attack}.
Developing an architecture capable of estimating the victim radar's parameters in real-time with sufficient accuracy presents unique technical challenges. Moreover, estimation errors can propagate throughout the rest of the attack implementation and impact the attack effectiveness. For example, if an attacker's estimate of the victim's frame start time is off by even \nsec{20}, a spoofing FP attack's perceived location can be off by \m{3} in the victim radar's view.

While~\cite{vennam_mmspoof_nodate} implemented a black-box FP attack by estimating a victim radar's \emph{chirp period} and \emph{chirp slope}, we enable FN and translation attacks by introducing a novel sensing architecture that additionally estimates the \emph{frame period} in \emph{real-time} while simultaneously predicting future radar \emph{frame start times}. We show that our design is 
sufficiently accurate to enable effective attacks -- %. For example, our full-scale simulation 
e.g., {\name} estimates a victim's chirp slope and period with a mean error of \MHzPerus{0.01} and \nsec{0.14}, respectively; these highly accurate estimates result in 90\% of spoofing attacks being within \m{1.09} and \mPers{0.12} of the desired range and velocity, respectively. Lastly, as our approach observes only six victim frames, the attacker can quickly learn a victim's parameters, making our spoofing attacks significantly more practical compared to the white-box attacks implemented by previous works~\cite{sun_who_2021,miura_low-cost_2019,komissarov_spoofing_2021,chauhan_platform_2014}.

% Note that, as previously mentioned, all prior attack works focused on adding fake points into a victim radar's point cloud. With that said, 
While all prior FMCW radar security analyses (i.e.,~\cite{sun_who_2021,miura_low-cost_2019,komissarov_spoofing_2021,chauhan_platform_2014,vennam_mmspoof_nodate}) solely focused on FP attacks, other (non-security) works have shown that FMCW radars can be adversely, yet intermittently,  affected by naturally-occurring interference including same slope, similar slope, and sweeping interference~\cite{amar_fmcw-fmcw_2021, alland_interference_2019, schipper_discussion_2014,kunert_mosarim_2010, kunert_eu_2012,pietsch_more_2011}. These forms of interference occur when chirps with the same, similar, or different slopes are received by a radar, and may be caused by self-interference or interference from other radars in the environment. 
We build on these ideas and leverage specific forms of
interference to design effective on-demand FN attacks. To the best of our knowledge, \emph{this is the first work to present 
FN and translation attacks that effectively `remove' or `move/translate' detections of existing objects in a victim radar's point cloud}.
We accomplish this by introducing very similar slope interference as part of the attack, using the estimated victim radar's parameters. 
Further, we show that by leveraging the estimated parameters of the victim radar, the proposed attacks can be designed to result in \emph{multiple} FP and FN object detections (and thus, multiple translated detections) in every execution frame.
As part of our analysis, we show how spoofing and intentional interference attacks propagate through a radar’s Range-Doppler, CFAR point-detection, and DBSCAN clustering stages.

We demonstrate the feasibility and applicability of {\name} by developing a proof-of-concept prototype using the USRP B210 software-defined radio (SDR) platform~\cite{ettus_research_usrp_nodate}. The developed attack platform estimates the victim’s parameters and then uses those estimates to launch the desired attacks with (multiple) FP, FN, and translation outcomes, all in real-time.
Through simulation and physical experimentation, we show that our black-box attacker can estimate the victim's parameters with sufficient accuracy to launch successful attacks over 95\% of the time. 
We perform comprehensive attack evaluation on real-world case studies using our prototype to demonstrate various attack outcomes -- i.e., single and multiple FP, FN, and/or translation attacks, as well as successful attacks on victims employing basic defenses such as parameter randomization. Additional resources, including case study videos, and case studies can be found at~\cite{Project_Website}.

Our contributions can be summarized as follows:
\begin{itemize}%[topsep=1.0px, itemsep=1.0pt]
    \item We introduce {\name}, a black-box attack framework for effective physical layer attacks on mmWave radars without prior knowledge of the victim radar's parameters (e.g., the chirp period and slope, and frame~duration);
    \item We enable new black-box attack types by improving upon existing methods for estimating victim parameters;
    \item We demonstrate that mmWave radars are vulnerable to \emph{false-negative} and \emph{translation} attacks that effectively `remove' or `move' detections of existing objects in the victim's point cloud, respectively;
    \item We demonstrate feasibility, and evaluate our attacks on multiple real-world case studies performed using a real-time implementation on the USRP B210 SDR platform.
\end{itemize}

This paper is organized as follows. 
Section~\ref{Preliminaries} overviews the FMCW radar signal processing pipeline.
Attack objectives and threat model are introduced in Section~\ref{Attack_Objectives}.
Section~\ref{Attack_Design} describes the attack framework, starting from estimating the victim radar's parameters, before showing how such estimates can be used to launch the attacks. 
Given the cost and hardware limitations of our real-time physical prototype, we first present results of rigorous % simulations to characterize and predict 
simulation-based performance evaluation of the parameter estimation module in Section~\ref{Victim_Parameter_Sensing} and the full-scale attacks in Section~\ref{Large_Scale_Evaluation}.
Section~\ref{Physical_Implementation} presents our physical prototype and results from real-world evaluations, before multiple real-world case studies are introduced in Section~\ref{Real_World_Case_Studies} to demonstrate the performance and feasibility of our novel framework in realistic scenarios.
Finally, framework limitations and potential defense mechanisms are discussed in Section~\ref{Discussion_and_Future_Work}, before providing concluding remarks in Section~\ref{09_Conclusion}.

%% file: 02_Preliminaries.tex
\section{Preliminaries: FMCW Radar Signal Processing}
\label{Preliminaries}

Radars employ radio waves for sensing, by transmitting a specifically constructed signal into the environment. The transmitted signal reflects off objects in the radar's field of view; the reflections are then received (and processed) by the radar's receiver (Rx). In particular, the received signal is used to determine the range, velocity, and relative angle of objects in the environment. The ability to detect an object's velocity in a single frame is unique to radars as other sensors (e.g., cameras)  can at most determine an object's range (e.g., with stereo cameras) and angle from a single image frame. 
% \footnote{Both cameras and Lidars can detect an object's velocity, but this measurement is less accurate and requires multiple frames.}.

\begin{figure*}[!t]
\centering
\includegraphics[width=1.998\columnwidth]{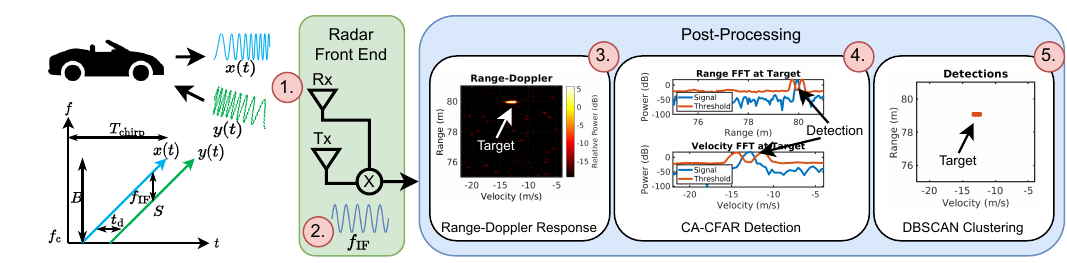} 
\caption{FMCW radar signal processing pipeline.}
\label{fig_FMCW_pipeline}
\end{figure*}

FMCW radars are a type of radar sensor commonly employed in automotive systems. They use a common signal~processing pipeline (Fig.~\ref{fig_FMCW_pipeline}) with the following five key steps.

%%%%%
%%%%%
\vspace{1.0ex}
\noindent\textbf{Step \circled{1}: Transmitter (Tx) and Rx chirps}.
In each frame, a radar transmits a series of identical ``\emph{chirps}'', whose frequency increases linearly over time. In general, a series of 256 chirps are transmitted per radar frame \cite{texas_instruments_iwr1443_2018}. Fig.~\ref{fig_FMCW_pipeline} illustrates the frequency response of a single Tx chirp and the corresponding Rx chirp, which is reflected off of an object in the environment and received by the radar. To capture radar parameters, which control sensing performance, we use the following notation: $\ChirpFreqStart$~denotes the \emph{chirp start frequency}, $\ChirpBW$ the \emph{chirp bandwidth}, $\ChirpDuration$ is the \emph{chirp period}, $\FreqIF$ the \emph{intermediate frequency} (IF) from mixing the Tx chirp with its corresponding Rx chirp, $\ChirpSlope$ is the \emph{chirp slope}, and $\LightSpeed$ denotes the \emph{speed of light}. 

Let $x(t)$ denote the FMCW radar Tx signal corresponding~to a single chirp in a radar frame, given by~\cite{alland_interference_2019,sun_who_2021,jiang_mmvib_2020, wang_remote_2020}\footnote{We use the common notation with {\eqref{eq_FMCW_tx_chirp}} expressing the transmitted signal at baseband where it has been sampled (in the digital domain) as a complex signal, composed of real and imaginary components, often denoted as the in-phase (I) and quadrature (Q) components. The two signals are identical, except that the Q signal is shifted by 90 degrees from the I component. The actual over-the-air radar signal is transmitted as a real-value analog signal.}
\begin{equation}
\label{eq_FMCW_tx_chirp}
x(t) = e^{ j \left(2 \pi \ChirpFreqStart \cdot t + \pi \ChirpSlope \cdot t^2 \right) }.
\end{equation}

Consider a target whose relative\footnote{Relative is with respect to the direction of propagation of the radar's Tx signal. Thus, relative range and velocity are scalars.} range and velocity at time $t$ are denoted by $d(t)$ and $\VelocityTarget$,\footnote{To simplify presentation, we assume constant velocity over the duration of a radar frame.} respectfully; thus, $d(t) = R(t) + \RangeTarget$, where $\RangeTarget$ is the target's initial position at the start of the radar frame and $R(t) = \int_{0}^{t} \VelocityTarget\ \textrm{d}t$ is the distance that it has traveled by time $t$~\cite{wang_remote_2020,jiang_mmvib_2020}.

As the signal propagates at the speed of light ($\LightSpeed$), the time $\tDelay$ it takes the radar signal to propagate to the target and back is given by $\tDelay = 2 d(t)/\LightSpeed$. Thus, the reflected signal received by the radar from the target, denoted by $y(t)$, can be captured as
\begin{equation}
\label{eq_FMCW_rx_chirp}
    y(t) = A_{\textrm{Rx}} \cdot e^{ j \left[2 \pi \ChirpFreqStart (t-\tDelay) + \pi \ChirpSlope (t - \tDelay)^2 \right] } + z(t),
\end{equation}
where $A_{\textrm{Rx}}$ and $z(t)$ denote the received signal amplitude and noise, respectively. 
% \footnote{For simplicity, we assume received signal power is constant over the duration of the frame}

%%%%%
%%%%%
\vspace{1.0ex}
\noindent\textbf{Step \circled{2}: Dechirping and IF signal generation}.
In this step, an IF signal is obtained by mixing the transmitted signal with the received signal; 
thus, the resulting signal $s_{\textrm{IF}}^{(l)}(t)$ corresponding to the $l$-th chirp is given~by
\begin{equation}
\label{eq_FMCW_IF_signal_simplified}
    s_{\textrm{IF}}^{(l)}(t) = x(t) \cdot y^{*}(t)
    = A_{\textrm{IF}} \cdot e^{j 2 \pi \FreqIF \cdot t} \cdot e^{j \DopplerShift l} + z'(t),
\end{equation}
where $\FreqIF := \frac{2 \ChirpSlope \cdot \RangeTarget}{c}$, $\DopplerShift := \frac{4 \pi \VelocityTarget \cdot \ChirpDuration}{\lambda}$, $\lambda = \LightSpeed/\ChirpFreqStart$ is the signal wavelength, $A_{\textrm{IF}}$ is the amplitude of the IF signal, and $z'(t) = x(t) \cdot  z^{*}(t)$ represents the noise present after the mixing ~\cite{wang_remote_2020,budge_range_1993} (for details see Appendix~\ref{Appendix_Preliminaries_Simplification_of_IF_Signal}).

While the transmitted and received chirp signals may have a bandwidth up to $\ChirpBW = 4\thinspace\textrm{GHz}$, modern automotive FMCW radars use a low-pass filter to remove all IF frequencies above \MHz{10--20}. For example, the TI IWR1443 mmWave FMCW radar has a maximum IF signal bandwidth of \MHz{15}~\cite{texas_instruments_iwr1443_2018}. The maximum IF frequency directly impacts the maximum range that a radar can detect objects at (as we show in {\eqref{eq_range_performance}}), as well as significantly reduces the cost of implementation. Also, as we show in Section \ref{Attack_Design}, this impacts the development of black-box attacks on radar.

%%%%%
%%%%%
\vspace{1.0ex}
\noindent\textbf{Step \circled{3}: Range-Doppler response}.
The IF frequency, $\FreqIF$, corresponding to a specific target is estimated using a fast Fourier transform (FFT) of the IF signal from~\eqref{eq_FMCW_IF_signal_simplified}; then, the target's range $\RangeTarget$ is computed using
\begin{equation}
\label{eq_range_computation}
    \RangeTarget = \frac{\FreqIF}{2\ChirpSlope} \cdot c.
\end{equation}
The range resolution $\RangeRes$, defined as the minimum required distance between two targets for a radar to distinguish them, and $\RangeMax$, the maximum detection range are defined as~\cite{rao_introduction_nodate} (details are provided in Appendix~\ref{Appendix_Preliminaries_Range_Resolution_and_Maximum_Range})
\begin{equation} \label{eq_range_performance}
    \RangeRes = \frac{\LightSpeed}{2 \ChirpBW},\hspace{10pt}
    \RangeMax = \frac{\FreqSamp \cdot \LightSpeed}{\ChirpBW},
\end{equation}
where $\FreqSamp$ is the radar's sampling rate of the IF signal.

Multiple chirps in a single frame can be used to detect the velocity of a target, leveraging 
% \footnote{The TI IWR1443 transmits up to 256 chirps per frame to detect velocity of objects \cite{texas_instruments_iwr1443_2018}}.
the slight phase shift, denoted by $\DopplerShift$, between chirps due to a target's relative velocity causing a small change in distance over a chirp's duration~\cite{rao_introduction_nodate}. $\DopplerShift$ can be estimated by taking an FFT across all chirps in a frame for each range bin; the resulting FFT will have a peak at $\DopplerShift$, and the relative velocity satisfies~\cite{rao_introduction_nodate}.
\begin{equation}
\label{eq_velocity_computation}
    v = \frac{\DopplerShift \cdot \lambda}{4 \pi \cdot \ChirpDuration}.
\end{equation}
In addition, the velocity resolution $\VelocityRes$ and maximum velocity $\VelocityMax$ follow (details provided in Appendix~\ref{Appendix_Preliminaries_Velocity_Resolution_and_Maximum_Velocity})
\begin{equation}
\label{eq_velocity_performance}
    \VelocityRes = \frac{\lambda}{2 \NumChirpsPerFrame \cdot \ChirpDuration},\hspace{10pt}
    \VelocityMax = \frac{\lambda}{4 \cdot \ChirpDuration},
\end{equation}
where $\NumChirpsPerFrame$ is the number of chirps in a radar frame. The Range-FFT and Doppler-FFT responses are often computed simultaneously using the 2D-FFT operation to generate the  \emph{Range-Doppler response} (as illustrated Fig.~\ref{fig_FMCW_pipeline}).

%%%%%
%%%%%
\vspace{1.0ex}
\noindent\textbf{Step \circled{4}: CFAR Detection.}
Constant false alarm rate (CFAR) detectors are commonly used to detect objects in the Range-Doppler response by estimating the relative noise levels around each Range-Doppler cell. In general, this is non-uniform as different objects in the radar's field of view may cause more clutter than other objects. Using the estimated noise level at each cell, the CFAR detector computes a threshold configured to achieve a specific probability of false alarm.~As~the~noise level is not constant, the threshold varies to account for the clutter %present 
in different regions of the Range-Doppler~response. 

A cell that has an amplitude above the computed threshold is classified as a detection~\cite{katzlberger_object_2018,rohling_radar_1983,rohling_ordered_2011}. 
Step \circled{4} in Fig.~\ref{fig_FMCW_pipeline} illustrates the computed CFAR detection threshold for the range and velocity domains of a normal target (note that the threshold is not constant). The two most widely used CFAR methods are continuous average CFAR (CA-CFAR) and ordered statistic CFAR (OS-CFAR). The probability of a CA-CFAR detector detecting an object significantly decreases in scenarios with abnormally high clutter in specific regions or with two closely located objects~\cite{katzlberger_object_2018,rohling_radar_1983}. %In this work, we 
In this work, we exploit this property to design FN events on systems employing CA-CFAR detectors, as these are more commonly used (e.g., in 
TI IWR1443 mmWave FMCW radar~\cite{texas_instruments_iwr1443_2018}), but the approach can also be extended to OS-CFAR detectors.

%%%%%
%%%%%
\vspace{1.0ex}
\noindent\textbf{Step \circled{5}: Clustering.}
The final step of the radar signal processing pipeline is to group the detection cloud points corresponding to the same object using a clustering algorithm. In this work, we focus on the commonly employed density-based spatial clustering of applications with noise (DBSCAN) algorithm~\cite{ester_density-based_1996}, which achieves two primary objectives:
(\emph{i})~identifying different targets in the radar's field of view by grouping together regions with a high density of detection points, and
(\emph{ii})~filtering out detection points corresponding to noise or multi-path reflections (see Step~\circled{5} in Fig.~\ref{fig_FMCW_pipeline}).

\subsection{Non-Adversarial Interference and Interference Mitigation}
\label{Related_Work_Interference}

Even without adversarial activity, radar interference may occur due to the increased proliferation of radar sensing in modern vehicles. FMCW radars are susceptible to three key types of interference: \emph{same slope, similar slope,} and \emph{sweeping interference} occurring when an interfering signal has a chirp slope that is the same, similar, or significantly different to the victim radar's chirp~\cite{amar_fmcw-fmcw_2021, alland_interference_2019}. Generally, interference is the result of multi-path reflections and non-malicious interference from another radar, and all forms of interference can degrade radar performance. Interference could saturate a victim radar’s Rx stage, decrease the signal-to-noise ratio (SNR) of perceived targets, and generate false peaks or ghost targets~\cite{kunert_mosarim_2010, kunert_eu_2012,pietsch_more_2011}, 
as well as impact a radar’s Range-Doppler response~\cite{kunert_mosarim_2010, kunert_eu_2012,pietsch_more_2011}. 
resulting in the radar losing a target altogether~\cite{alland_interference_2019, schipper_discussion_2014}.

\iffalse 
\vspace{0.25ex}
\noindent\textbf{Interference mitigation}.
Several mitigation techniques have been proposed, mainly targeting sweeping interference -- i.e., interference from chirps with greatly different slopes.  \cite{alland_interference_2019, kunert_mosarim_2010, kunert_d15_2010, barjenbruch_method_2015, bechter_automotive_2015, bechter_automotive_2017} proposed various techniques to detect the interference in the time domain of the IF signal and then repair the received signal by nullifying the interference. Other techniques have also been proposed to combat general interference including the use of specific polarization and frequency bands, randomizing chirp timing, and adding random phase shifts to each chirp~\cite{amar_fmcw-fmcw_2021,kunert_mosarim_2010,kunert_d15_2010}.
\fi

\vspace{1.0ex}
\noindent\textbf{Attacks using interference.}
To the best of our knowledge, no work has considered the effects that a malicious actor could have if they intentionally interfered with a radar using attacks based on carefully-crafted similar slope interference. Existing methods for mitigating (e.g., similar-slope) interference (e.g.,~\cite{alland_interference_2019, kunert_mosarim_2010, kunert_d15_2010, barjenbruch_method_2015, bechter_automotive_2015, bechter_automotive_2017})
%\todo{are these ok references?}
are developed under the assumption that the interference is only \emph{sporadic} and \emph{not adversarial} (i.e., malicious). 
In general most mitigation techniques detect the interference in the time domain of the IF signal and then repair the received signal by nullifying the interference.
In this work, we also introduce attacks based on very similar slope interference that result in a FN event for victims employing a CA-CFAR detector. Our attacks are not detectable in the time domain, therefore making such interference mitigation techniques ineffective.

%% file: 03_Attack_Objectives.tex
\section{Attack Objectives and Threat Model} \label{Attack_Objectives}

\begin{figure}[!t]
    \centering
    \includegraphics[width=0.998\columnwidth]{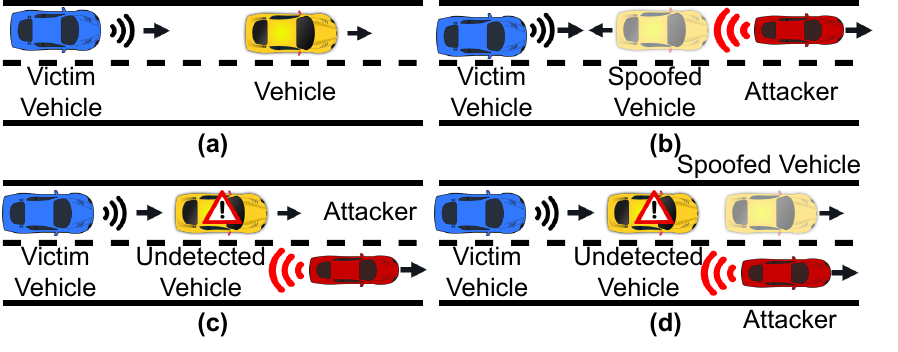}
    % \vspace{-4pt}
    \caption{Example traffic scenarios for (a) ao attack, (b) false positive attack, (c) false negative attack, and (d) translation attack (graphics from \cite{blue_car,red_car,yellow_car,wireless_signal,danger_sign}).}
    \label{fig_threat_model}
\end{figure}

We consider %the threat model depicted 
representative attack scenarios illustrated in Fig.~\ref{fig_threat_model}. Here, we refer to the \emph{victim} vehicle as the vehicle performing normal radar sensing operations. The \emph{attacker}'s goal is to produce incorrect sensing outcomes for the victim. The attacker may wish to orchestrate attacks in some relation to an existing \emph{target} object (e.g., another vehicle) other than the victim, in order to compromise safe victim vehicle operation.

\subsection{Attack Strategy and Goals} 
\label{Attack_Objectives_Attack_Strategy_and_Goals}
We introduce the \emph{false positive (FP), false negative (FN),} and \emph{translation} attacks that % our attacks 
use specifically designed signals to \emph{add} fake targets (FP), \emph{remove} real targets~(FN), or manipulate the range and velocity of existing targets (translation).

\vspace{1.0ex}
\noindent\textbf{FP attack.}
The first considered attack goal is to cause a FP sensing outcome, where an attacker inserts a spoofed (i.e., `fake') object into the victim radar's point cloud as illustrated in Fig.~\ref{fig_threat_model}(b). This is consistent with the existing FP outcomes for camera/{\lidar} attacks (e.g.,~\cite{sun_towards_2020,abdelfattah_adversarial_2021,hallyburton_security_2022,cao_adversarial_2019}). Intuitively, to achieve this the attacker should send (slightly delayed) chirps identical to that of the victim, emulating the signal reflected from a (spoofed) object. However, unlike attacks on camera and {\lidar} where FP attacks only aim to add a spoofed object at a particular range (i.e., distance) from the victim, with radar attacks, the goal is to spoof \textbf{\emph{both an object's range and velocity}}. As such, spoofed objects must update their position in consecutive frames based on the desired spoofing velocity; this allows the attack to propagate from perception to the tracking and prediction modules in autonomous vehicles.

\vspace{1.0ex}
\noindent\textbf{FN attack.}
The second %considered 
attack goal is to cause a FN sensing outcome, where the victim fails to perceive an existing physical object (Fig.~\ref{fig_threat_model}(c)). While a FN outcome is hard to achieve with black-box attacks on camera or {\lidar}-based sensing, intuitively our approach is to transmit an attacking signal that adds clutter around a desired target, therefore significantly lowering the CA-CFAR detection probability of the object.
% As we show in Sec.~\ref{Large_Scale_Evaluation}, this often leads to a false-negative event.

\vspace{1.0ex}
\noindent\textbf{Translation attack.}
The final commonly considered attack goal is to cause a translation event %where an attacker 
that effectively `moves' a real object from the victim's point of view (Fig.~\ref{fig_threat_model}(d)). This is achieved by launching a FN attack to `remove' an actual target while simultaneously employing a FP attack to `insert' a fake object into the victim's point cloud; %The result is that 
as result, the victim fails to detect the real object but detects the fake object.

\subsection{Environmental Assumptions} 
\label{Attack_Objectives_Environmental_Assumptions}

We make the following three assumptions:
\begin{enumerate}
\item[(\emph{i})]
The victim employs the radar processing pipeline from Fig.~\ref{fig_FMCW_pipeline}. While we focus on the %particular, 
most commonly employed radar sensing pipeline (e.g., 
TI IWR1443 mmWave FMCW radar~\cite{texas_instruments_iwr1443_2018}), the presented security analysis and attacks are generalizable to other %approaches based on a 
similar radar designs;
%\todo{check}
%%
\item[(\emph{ii})]
We focus on attacking only the victim's FMCW radar sensor. While most vehicles feature additional sensors (e.g., cameras), our objective is to show that \name's novel \emph{black-box} attacks are feasible~and~effective;%\todo{check}
\item[(\emph{iii})]
We focus on attacking only in the range and velocity domains for the initial black-box attack development. Thus, we assume that the attacker is physically located at the desired angle of attack; e.g., Fig.~\ref{fig_threat_model}(b) shows the case where the attacker is in front of the victim.
\end{enumerate}

\subsection{Attacker Capability and Knowledge}
\label{Attack_Objectives_Attacker_Capability_and_Knowledge}

We consider physical spoofing attacks, where the attacker can only transmit signals in order to achieve the desired attack outcomes. Unlike existing work, we consider the \emph{black-box} threat model where the attacker has no knowledge of the radar parameters utilized by the victim. We do assume that the attacker has knowledge about the environment; in particular, the victim's relative position and velocity for FP attacks so that a spoofed object behaves like a realistic target, as well as the position and velocity of the target object for FN attacks.

%% file: 04_Attack_Design.tex
\section{\name~Attack Design}
\label{Attack_Design}
We now present the methodology used to estimate the victim radar's parameters and show how such estimates can be used to launch FP, FN, and translation attacks. Fig.~\ref{fig_attack_architecture} overviews the design of {\name}'s black-box attack generator.

\subsection{Parameter Estimation} 
\label{Attack_Design_Parameter_Estimation}

Black-box attacks present a particularly difficult challenge as it is critical that the attacker can accurately estimate, in real-time, the key victim radar parameters. While the architecture from \cite{vennam_mmspoof_nodate} developed a black-box FP attack by estimating a victim's chirp period ($\ChirpDuration$) and chirp slope ($\ChirpSlope$), we need to additionally estimate a victim's frame duration ($\FrameDuration$) and predict future frames to develop our novel black-box FN and translation attacks. This presented unique technical challenges as accurate attacks require very precise predictions for when the next victim frame will occur. For example, if the prediction is off by \nsec{20}, the victim will perceive the spoofed object to be \m{3} away from where the attacker intended to add an object. 
Specifically, the resulting range error satisfies
\begin{equation}
    \label{eq_spoof_error}
    \RangeError = \LightSpeed \cdot \TimingError / 2,
\end{equation}
where $\TimingError$ is the frame start time prediction error.

\begin{figure}[!t]
    \centering
    \includegraphics[width=0.998\columnwidth]{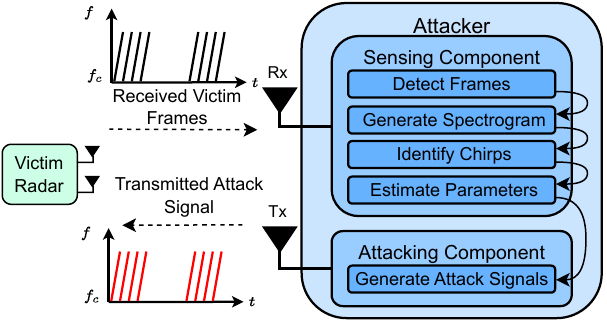}
    % \vspace{-12pt}
    \caption{\name~block diagram.}
    \label{fig_attack_architecture}
\end{figure}

We now introduce a real-time sensing module that enables effective FP, FN, and translation attacks by estimating the victim radar's parameters with low estimation errors; this is performed using three key steps summarized in Fig.~\ref{fig_attack_architecture}. 

\vspace{1.0ex}
\noindent\textbf{Step 1: Spectrogram generation.}
Similar to \cite{vennam_mmspoof_nodate}, we start by generating a spectrogram for each detected victim frame. The \name~prototype runs at {25}\thinspace{MSps} sampling rate and checks for victim frames every \msec{0.16}. Once a frame is detected
by a custom frame detector that tracks increases in received~power, we record the received signal for %a duration of just 
slightly over \msec{2}, and then generate a spectrogram that samples the frequency every \usec{2}\cite{proakis_design_2007,ayguen_pocketfft_nodate}. This computation is done in under~\msec{10}.%\todo{check}

%\noindent\textbf{Step 1: Frame start detection}.
%The start of a victim radar frame is detected when the received signal power at the attacker exceeds its Rx's noise floor by a given threshold, which is empirically set to be {5}\thinspace{dB}. The Rx continuously captures the received signal, computes the signal power in real-time, and determines if a victim frame is detected. In our prototype, the attacker runs at {25}\thinspace{MSps} sampling rate and checks for frames every {0.16}\thinspace{ms} (see Section~\ref{Physical_Implementation}). 

%\vspace{0.25ex}
%\noindent\textbf{Step 2: Spectrogram generation}.
%Once a victim radar frame is detected, the sensing component records the received signal for a duration of just over {2}\thinspace{ms}, which corresponds to about 51,000 samples at {25}\thinspace{MSps} sampling rate. Next, a Hanning window~\cite{proakis_design_2007} is applied to small windows over the received signal, followed by the FFT for each window. Finally, we generate a spectrogram by taking the log-amplitude of each window's FFT. In \name's~prototype, we estimated the frequency every \usec{2} using a 50 sample buffer. From there, we leverage the C++ code from~\cite{ayguen_pocketfft_nodate} to compute 32-bin FFTs on the first 32 samples of each buffer to estimate the frequency. In total, 1,020 FFTs are computed to estimate the frequency every \usec{2} for the full \msec{2} recorded signal. The \name~prototype performs the full computation in under~\msec{10}.

\vspace{1.0ex}
\noindent\textbf{Step 2: Identify chirps in spectrograms.}
While \cite{vennam_mmspoof_nodate} used signal energy over time to estimate the chirp period ($\EstimatedChirpDuration$) and the spectrogram of a single chirp to estimate the chirp slope ($\EstimatedChirpSlope$), we designed and implemented a peak detection and clustering algorithm to identify the (time, frequency) points corresponding to each chirp within the generated spectrogram. For each chirp's  (time, frequency) points, we use least squares regression to estimate the
%$\widetilde{\textbf{b}}_{i} := [\widetilde{b}_{0,i}, \widetilde{b}_{1,i}]^{\top}$, where $\widetilde{b}_{0,i}$ and $\widetilde{b}_{1,i}$ denote the estimated 
start time and slope for the $i$-th detected chirp, respectively~\cite{chamberland_engineering_2020}.
The estimates are computed in real-time using the Eigen C++ library~\cite{noauthor_eigen_nodate}.

\vspace{1.0ex}
\noindent\textbf{Step 3: Estimate victim parameters.} 
Accurate estimates of the victim radar's chirp slope ($\EstimatedChirpSlope$), chirp period ($\EstimatedChirpDuration$), and frame duration ($\EstimatedFrameDuration$), are achieved by averaging their computed values across multiple recorded victim chirps and frames. While averaging over a large number of computed parameters results in sufficiently accurate estimates for the chirp period and chirp slope, we note that we collect far fewer samples for the frame period (e.g., up to 256 chirps per frame). Thus, we use cross-correlation to compute a more precise frame start time. Specifically, we take the cross-correlation of the first \usec{10} of the received signal and a computed victim chirp (generated using the estimated parameters) to further improve the accuracy of the estimated frame start time. As shown in our experiments and simulations (Section~\ref{Victim_Parameter_Sensing}), %using the cross-correlation generated 
this results in sufficiently accurate estimates after only 6 frames.
%\todo{check}

In rare cases, the implemented sensing component experiences errors (e.g., classifying one chirp as two different chirps) resulting in significantly incorrect measurements. We account for these erroneous estimates by using the inter-quartile range to identify and filter out outliers~\cite{degroot_probability_2023}. The end result is a robust sensing component capable of quickly and accurately estimating a victim radar's parameters.

\subsection{False Positive Spoofing Attacks}
\label{04_false_positive_spoofing}

\begin{figure}[!t]
    \centering
    \includegraphics[width=0.996\columnwidth]{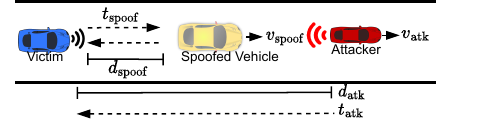}
    \caption{Key attack parameters.}
    \label{fig_attack_design}
\end{figure}
\begin{figure}[!t]
    \centering\includegraphics[width=0.998\columnwidth]{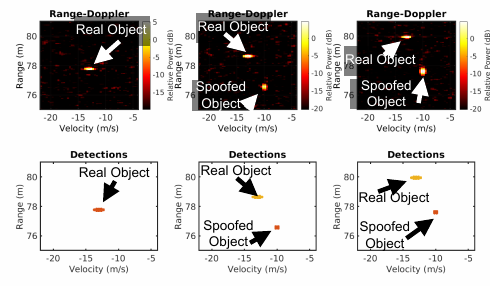}
    % \vspace{-6pt}
    \caption{FP attack at frames 1 (no attack), 7, and 10, adding a false moving object in addition to the real moving vehicle. Range-Doppler Plot (Step~3 in Fig.~\ref{fig_FMCW_pipeline}) shown in top row;~Identified Clusters (Step 5 in Fig.~\ref{fig_FMCW_pipeline}) are shown in the bottom row. }\label{fig_simulated_false_positive_attack_progression}
\end{figure}
Intuitively, the attacker uses the \emph{estimated} chirp slope ($\EstimatedChirpSlope$), chirp period ($\EstimatedChirpDuration$), and frame duration ($\EstimatedFrameDuration$) to launch a FP spoofing attack by transmitting identically sloped radar chirps with a specific delay, $\AttacktDelay$, and phase shift, $\AttackDopplerShift$. Fig.~\ref{fig_attack_design} summarizes the key parameters used when constructing the FP attack. Unlike existing white-box FP radar attacks~\cite{sun_who_2021,komissarov_spoofing_2021}, \name's~black-box attack framework \emph{does not} assume a-priori knowledge of the victim radar's parameters; % of the victim radar a-priori.Specifically, we design 
it rather uses $\AttacktDelay$ and $\AttackDopplerShift$ based on the desired position and velocity of the spoofed object, respectively, obtained as 
\begin{equation}
\begin{split} 
\label{eq_attack_delay_and_doppler_shift}
    \AttacktDelay &= t_{\textrm{spoof}} - t_{\textrm{atk}} 
        = \frac{1}{\LightSpeed} \cdot \left(2\RangeSpoof - \RangeAttacker \right), \\
    \AttackDopplerShift &= \frac{4\pi}{\lambda} \cdot  \left(\VelocitySpoof - \frac{\VelocityAttacker}{2}\right) \cdot \EstimatedChirpDuration \cdot n,
    \end{split}
\end{equation}
where $\RangeSpoof$ is the desired spoofing range, $\RangeAttacker$~is~the relative range of the victim w.r.t the attacker, $t_{\textrm{spoof}}$ is the time delay corresponding to a target at $\RangeSpoof$, $t_{atk}$ is the propagation delay for a signal to travel $\RangeAttacker$ from the attacker to the victim, 
$\VelocitySpoof$ is the desired spoofing velocity, $\VelocityAttacker$ is the relative velocity of the victim (w.r.t the attacker), and $n$ is the attack chirp index. 

Here, we emphasize the importance of accurate estimation of the victim's position and velocity, % (see Section~\ref{Attack_Objectives}), 
which allows for the attacker to spoof objects at specific positions and velocities. %To further improve the realism of our attacks, we 
We also dynamically scale the amplitude of the Tx signal, denoted by $\AttackAmplitude$, to emulate the propagation loss that scales with $4 \pi \RangeSpoof^{2}$. Based on {\eqref{eq_attack_delay_and_doppler_shift}} and the obtained parameter estimates, the $n$-th chirp of the FP attack signal is computed~by
\begin{equation}
\label{eq_FP_attack_chirp}
    x^{(n)}_{\textrm{FP}}(t) = \AttackAmplitude \cdot e^{j\left[2 \pi \ChirpFreqStart(t-\AttacktDelay) + \pi \EstimatedChirpSlope (t-\AttacktDelay)^{2} + \AttackDopplerShift\right]}.
\end{equation}
Fig.~\ref{fig_simulated_false_positive_attack_progression} shows the Range-Doppler response and point cloud over multiple frames for a simulated FP attack. The ``real object'' in the scene is the attacker while the ``spoofed object'' is a fake object that the attacker intends to insert. Note that the spoofed object exhibits realistic motion and has a power level expected for objects at that range. 

\vspace{15pt}
\subsection{False Negative Attacks}
\label{04_alse_negative_attacks}

\begin{figure*}[!t]
    \centering
    \includegraphics[width=0.998\textwidth]{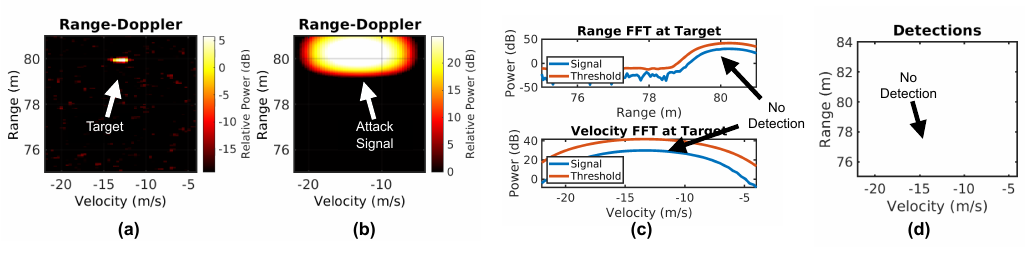}
    % \vspace{-10pt}
    \caption{An example of a FN Attack: (a)~Range Doppler plot without attack (step 3 in Fig.~\ref{fig_FMCW_pipeline}); (b)~Range Doppler plot (step 3 in Fig.~\ref{fig_FMCW_pipeline}); (c)~CFAR threshold (step 4 in Fig.~\ref{fig_FMCW_pipeline}); and (d)~resulting in no identified clusters (step 5 in Fig.~\ref{fig_FMCW_pipeline}).} \label{fig_simulated_FN_key_processing_steps}
\end{figure*}

Intuitively, \name~attacks achieve a FN outcome by adding clutter in the Range-Doppler response around a specific target, in order to raise the CA-CFAR detection threshold and thus %. Thus, we 
significantly decrease the probability of the actual target being detected. 
We start with the FP attack signal from~\eqref{eq_FP_attack_chirp} that spoofs a false object at the same range and velocity as an actual target. Next, we slightly smear the spoofed signal in the range domain by using a very similar slope ($\AttackSlopeFN$) that is slightly offset ($\sim$\MHzPerus{0.01}) from the estimated victim slope ($\EstimatedChirpSlope$). This offset is computed to smear the spoofed signal by an additional {1--3}\thinspace{m} in the range domain and accounts for the victim radar's estimated bandwidth, chirp period ($\EstimatedChirpDuration$), and chirp slope ($\EstimatedChirpSlope$).
Finally, the spoofed signal is smeared in the velocity domain by subtly increasing the $\AttackDopplerShift$ phase shift between subsequent chirps; the phase shift for each chirp~is
%\todo{check}
%
\begin{equation*} %\label{eq_sim_vel_attack_signal}
        \AttackDopplerShiftFN{n+1} = \frac{4 \pi [( v_{0}+n\cdot\Delta\phi') - \VelocityAttacker/2] \EstimatedChirpDuration}{\lambda} + \AttackDopplerShiftFN{n}, 
\end{equation*}
where $n$ is the attack chirp index, $\Delta \phi'$ is the amount that the $\phi_{doppler}$ increases with each chirp, and $v_{0}$ is a velocity slightly less than $\VelocitySpoof$ so that the added clutter is centered at $\VelocitySpoof$.

Thus, the $n$-th chirp of the FN attack signal is given by
\begin{equation}
\label{eq_FN_attack_chirp}
    x^{(n)}_{\textrm{FN}}(t) = \AttackAmplitude \cdot e^{j\left[2 \pi \ChirpFreqStart(t-\AttacktDelay) + \pi \AttackSlopeFN (t-\AttacktDelay)^{2} + \AttackDopplerShiftFN{n}\right]}.
\end{equation}
The resulting attack adds clutter specifically around the target in a way that the CA-CFAR 
fails to detect the object, as shown in Fig.~\ref{fig_simulated_FN_key_processing_steps}.
In particular, the added clutter increases the CFAR threshold in the range and velocity domains at the target location to the point that no object is detected (see Fig.~\ref{fig_simulated_FN_key_processing_steps}(c)).
%In particular, Fig.~\ref{fig_simulated_FN_key_processing_steps}c presents the CFAR threshold in the range and velocity domains at the target location -- the threshold has been raised to the point that no object is detected. 

While the FN attack signal shown in Fig.~\ref{fig_simulated_FN_key_processing_steps}(b) appears easy to identify in the Range-Doppler response, it is unlikely to be detected on current FMCW radars. Most automotive radar systems only utilize CFAR detectors to detect objects in the Range-Doppler response followed by a clustering algorithm to group detections from the CFAR detector. Thus, a real object stays undetected when an attack causes a FN event in the CA-CFAR detector \cite{katzlberger_object_2018,rohling_radar_1983,rohling_ordered_2011}. Additionally, the added clutter is localized around the Range-Doppler bin corresponding to a specific target such that the overall noise level of the entire Range-Doppler response is only slightly  raised.~Therefore, it is unlikely that our attacks would be detected by additional monitoring of the noise level of the overall Range-Doppler response. Moreover, most existing interference mitigation methods (e.g.,~\cite{alland_interference_2019,kunert_eu_2012}) would not be able to detect the attack as the IF signal for the FN attack appears identical to the IF signal from a normal target. Finally, while it would be possible to design an algorithm to detect the added clutter in the Range-Doppler response (e.g., via DNNs), %such systems 
their use would require high computation costs and we are unaware of any such algorithms are currently implemented on commercial~systems.

\subsection{Translation Attack}
The translation attack is achieved by simultaneously transmitting the FP attack from {\eqref{eq_FP_attack_chirp}} and the FN attack from \eqref{eq_FN_attack_chirp}. For the FN attack, we set $\RangeSpoof$ and $\VelocitySpoof$ to the location of an actual target so that it is 'removed' from the victim radar's point cloud. For the FP attack,  $\RangeSpoof$ and $\VelocitySpoof$ are set to the location where we want the victim to detect the object. The result of the combined FP and FN attack is that an object~in the victim radar's point cloud is 'moved' as the attacker desires.

%% file: 05_Victim_Parameter_Sensing.tex
\section{Evaluation of Victim Parameter Estimation} 
\label{Victim_Parameter_Sensing}
%
%\tingjun{In general, describe things in present tense, we perform... we consider... we utilize...}
%

In Section~\ref{Physical_Implementation}, we present a real-time {\name} physical prototype  %of our attack 
 %framework 
developed using %software-defined radios, 
SDR platforms, but we were limited by the available hardware. Thus, we first employ rigorous simulations to emulate real-world conditions and predict the performance of {\name}'s framework on a full-scale system. In this section, we present an evaluation of the sensing module before evaluating the full attack %accuracy and 
performance (Section~\ref{Large_Scale_Evaluation}).
%\todo{check}

%%%%%
%%%%%
\subsection{Simulation Environment and Setup}
\label{sec:sim_env}

We generated realistic environments with multiple objects utilizing the Matlab Phased Array System Toolbox~\cite{noauthor_phased_nodate}. To start, we used the toolbox's {\small\texttt{RadarTarget}} object to simulate the behavior of radar signals reflecting off of moving targets. Each target's radar cross-section is randomly selected using a normal distribution with a mean of \dBsm{15} and a variance of \dBsmSq{5}, corresponding to the cross-section of a common midsized vehicle~\cite{alland_interference_2019}. Next, we simulate the effects of range-dependent time delays, propagation losses, phase shifts, and doppler shifts due to signal propagation using the toolbox's {\small\texttt{FreeSpace}} object, where the environment thermal noise level is given by $-174 + 10 \log_{10}\ChirpBW$ (dBm). The Tx's and Rx's in our radar and attacker implementations are simulated using objects from the toolbox's {\small\texttt{Transmitter}} and {\small\texttt{ReceiverPreamp}}, respectively. Specifically, each Tx has a Tx gain of \dB{36} and an output power of \dBm{5} without the Tx gain, and each Rx has an Rx gain of {42}\thinspace{dB} and a noise figure of {5}\thinspace{dB}~\cite{mathworks_radar_nodate}. Thus, the simulated environment features realistic Tx's, Rx's, targets, and signal propagation effects. Finally, we utilize the  {\small\texttt{RangeDopplerResponse}} and {\small\texttt{CFARDetector}} objects from MATLAB's Phased Array Toolbox and {\small\texttt{dbscan}} clustering algorithm from MATLAB's Machine Learning Toolbox to implement the victim radar signal processing pipeline~\cite{noauthor_ML_nodate}.

%%%%%
%%%%%
\vspace{1.0ex}
\noindent\textbf{Experimental setup.}
We evaluate the sensing module using the simulated environment with 200 different victim configurations based on realistic parameters of the TI IWR1443 mmWave FMCW radar~\cite{texas_instruments_iwr1443_2018}. Specifically, the victim radar's chirp slope is sampled uniformly at random from the interval [\MHzPerus{1}, \MHzPerus{100}],
%\tingjun{uniformly at random? what type of random distribution?}
%\david{clarified distribution}
and the chirp period is uniformly sampled %at random 
from the interval [\usec{15}, \usec{100}]. %\footnote{The lower bound was set to $15 \mu s$ as we found sensing module required at least $15 \mu s$ worth of samples to accurately estimate a chirp's slope and start time.} 
%\tingjun{if you are doing this why don't just go with the smallest chirp duration of 30us and longest chirp of 35 us???}
%\david{Under the slope (1MHz/us to 100MHz/us), and bandwidth (30 MHz to 3.5 GHz) constraints, shortest chirp would be (S = 100 MHz/us, B = 30 MHz) 0.3 us, longest chirp would be (S = 1MHz/us, B = 3.5 GHz) 3.5 ms. Given this, I decided to constrain the chirp period to be 15us to 100us so as to have a range of more reasonable chirp periods.}
To encompass the majority or radar configurations found in the automotive domain, the chirp bandwidth is imposed to be within [{30}\thinspace{MHz}, {3.5}\thinspace{GHz}]. Finally, we set a radar frame rate of {33}\thinspace{Hz}, which is the maximum frame rate of an automotive radar ~\cite{cheng_person_2022,tilly_detection_2020}. More details on the test cases can be found in Appendix~\ref{Appendix_Simulation_Details_Sensing_Component_Testing}.

For each victim configuration, 7 victim frames are simulated. We use the estimated victim frame duration ($\EstimatedFrameDuration$), chirp period ($\EstimatedChirpDuration$), and chirp slope ($\EstimatedChirpSlope$) at the end of the 7-th frame to assess the parameter estimation accuracy achieved by {\name}. We also record the predicted frame start time for each victim frame to understand how the estimation accuracy changes with an increased number of detected victim frames. Finally, %we verify that 
to evaluate {\name}'s parameter estimation %component functions 
accuracy %of the victim's parameters 
regardless of the victim position and velocity, we
uniformly sample the attacker's relative (w.r.t the victim) range ($\RangeAttacker$) and velocity ($\VelocityAttacker$) at random from the interval [\m{20}, \m{100}] and [\mPers{$-$10}, \mPers{10}].
%selecting the attacker's relative range and velocity from the interval [\m{20},\m{100}] and [\mPers{-20},\mPers{20}] uniformly at random, respectively.
%\todo{check}
%\tingjun{did you emulate the process that the victim sends out chirps which get bounced off of environment objects, etc. (i.e., are you really simulating the propagation of the signal or just numerically evaluating the equations?)}
%\david{Addressed in "simulation environment" section}
%\tingjun{so where does the uncertainty come from? this description of the simulation setup sounds to me like you will just get the absolute correct timing since there is no "noise"} 
%\david{ Addressed in "simulation environment" section}

%%%%%
%%%%%
\subsection{Simulation Results}
We now present the results from our simulated evaluations. 
We quantify how estimation errors for the victim radar's parameters lead to spreading 
%$\RangeSpread$ from~\eqref{eq_range_FFT_spread}
and spoofing  ($\RangeError$ from~\eqref{eq_spoof_error})  errors.
%\todo{check} 
Note that the spoofing velocity is generally unaffected as timing and slope estimation errors almost solely affect the range spoofing performance. 
%\tingjun{It is very unclear to me, how do you test and evaluation the parameter sensing accuracy IN SIMULATION? If you manually inserted noise and other artifacts then you need to describe them. }
%\david{In the previous section, I've added two sentences stating how we inserted noise and the victim/attacker transmit power, receive gain, and receive noise figure. Additionally, over the simulation run, some configurations ended up being more difficult for the radar to estimate parameters for than other configurations. TODO: add further clarification if needed}

\begin{figure}[!t]
    \centering    \includegraphics[width=0.998\columnwidth]{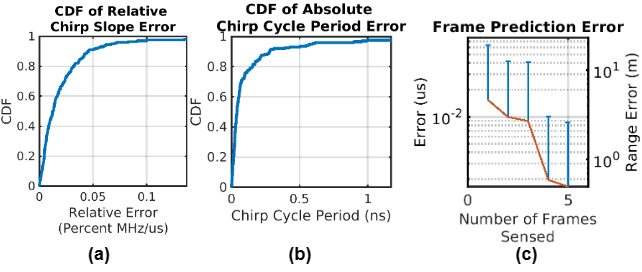}
    % \vspace{-6pt}
    \caption{Distributions for parameter estimation errors.}\label{fig_simulation_sensing_error_cdfs}
    % \vspace{-2pt}
\end{figure}

\vspace{1.0ex}
\noindent\textbf{Estimation accuracy for victim chirp slope and period.}
Figs.~\ref{fig_simulation_sensing_error_cdfs}(a) and~\ref{fig_simulation_sensing_error_cdfs}(b) show the cumulative density function (CDF) of the absolute and relative estimation errors for the victim radar's chirp period and chirp slope, whose key metrics are summarized in Table~\ref{table_simulation_parameter_estimation_results}.
The results show that 95\% of the estimated chirp period values, $\EstimatedChirpDuration$, are within \nsec{0.59} of their actual values, which corresponds to less than \m{0.1} of the spoofing error, $\RangeError$, based on {\eqref{eq_spoof_error}}. Also, %the results show that 
95\% of the estimated chirp slope values, $\EstimatedChirpSlope$, are within 0.03\% of their actual values. While this error could result in some smearing in the range domain, we show in Section~\ref{Large_Scale_Evaluation} that this is sufficient enough to launch successful attacks.

\begin{table}[!t]
  \caption{Key parameter estimation error metrics.}\label{table_simulation_parameter_estimation_results}
  % \vspace{-6pt}
  \centering
    \begin{tabular}{cccl}
        \toprule
        Parameter & Error Type & Mean Error & 95-th Percentile\\
        \midrule
        Chirp Period & Absolute & \nsec{0.143} & \nsec{0.586} \\
        Chirp Slope & Absolute & \MHzPerus{0.010} & \MHzPerus{0.0354}\\
         Chirp Slope & Relative & $0.025 \%$ & $0.068 \%$ \\
        \bottomrule
    \end{tabular}
    % \vspace{-4pt}
\end{table}

\vspace{1.0ex}
\noindent\textbf{Accuracy vs. Number of measured frames}.
Fig.~\ref{fig_simulation_sensing_error_cdfs}(c) plots the average absolute error for the predicted frame start times for each of the 200 frames, where the error bars represent the 95-th percentile of the absolute error. The left y-axis reports the prediction error in $\upmu$s while the right y-axis reports the resulting spoofing error in meters using \eqref{eq_spoof_error}. % from Section~\ref{Attack_Design}. 
%\tingjun{how should I read the double y-axis of 6c? the plot is not clear}
%\david{addressed and pointing to eq(8)}
It can be seen that the 
prediction error decreases as the number of number of considered frames increases. Further, the absolute error significantly decreases after the third victim frame is detected when the sensing component begins using the cross-correlation and the computed victim chirp (from the estimated parameters) to achieve more accurate frame start-time estimates.  Overall, in 95\% of cases, {\name} sensing can predict a victim radar's next frame start time with an accuracy that corresponds to less than \m{2} of range spoofing error ($\RangeError$) based on {\eqref{eq_spoof_error}}.

%% file: 06_Large_Scale_Evaluation.tex
\section{Large Scale Evaluations}
\label{Large_Scale_Evaluation}

The simulation environment introduced in Section~\ref{Victim_Parameter_Sensing} was used for %  we now present 
several large-scale evaluations %establishing 
of the accuracy and effectiveness of the \name~black-box attack framework. Specifically, we performed  \emph{5,000 unique simulations} to demonstrate %that our attack architecture launches successful and accurate 
the effectiveness and accuracy of the attacks regardless of the relative range and velocity of the victim. We considered the four victim radar configurations from Table~\ref{table_simulation_spoofing_victim_configurations} covering
a broad assortment of %potential 
automotive radar configurations.  Note that $\RangeMin$ represents the minimum CFAR detection range for each configuration (the full CFAR detection region is provided in Appendix~\ref{Appendix_Simulation_CFAR_Detection_Region}). Here, we start by evaluating spoofing accuracy using configurations C and D, % as these are 
representative of typical automotive long range radar (LRR) and short range radar (SRR), respectively. Then we evaluate how FP and FN attacks effect a victim's \emph{probability of false alarm} (PFA) and \emph{probability of detection} (PD) across all four configurations.
%\todo{check} 
%\tingjun{I am confused: do we use config A and B at all? or only C and D??}
%\david{Addressed in the last two sentences}

\begin{table}[!t]
\begin{center}
    \caption{Victim configurations for attack evaluation. 
    %\tingjun{update parameter notation}
    %\david{updated}
    }\label{table_simulation_spoofing_victim_configurations}
    % \vspace{-6pt}
    \begin{tabular}{ 
    m{.12\columnwidth}
    m{.12\columnwidth}|
    m{.12\columnwidth}
    m{.12\columnwidth}
    m{.12\columnwidth}
    m{.12\columnwidth}}
    \toprule
    Parameter & (unit) & A &  B &  C &  D \\
    \midrule
    $\ChirpFreqStart$ & GHz & 77.0 & 77.0 & 77.0 & 77.0 \\
    $\ChirpBW$ & MHz & 27.81 & 96.31 & 1001.51 & 3935 \\
    $\ChirpSlope$ & MHz/us & 1.15 & 4.04 & 47.85 & 187.96 \\
    $\ChirpDuration$ & us & 24.11 & 23.85 & 20.93 & 21.63 \\
    $\NumChirpsPerFrame$ & & 128 & 256 & 256 & 256 \\
    $\RangeRes$ & m & 7.49 & 2.14 & 0.21 & 0.05 \\
    $\RangeMax$ & m & 479 & 273.93 & 219.27 & 223.22 \\
    $\RangeMin$ & m & 44.97 & 19.26 & 2.14 & 0.54 \\
    $\VelocityRes$ & m/s & 0.62 & 0.31 & 0.35 & 0.35 \\
    $\VelocityMax$ & m/s & 39.49 & 39.91 & 45.00 & 44.99 \\
    \bottomrule
    \end{tabular}
\end{center}
% \vspace{-4pt}
\end{table}
%

%%%%%
%%%%%
\subsection{Spoofing Accuracy Evaluation}
\label{Large_Scale_Evaluation_Spoofing_Accuracy_Evaluation}

\noindent\textbf{Setup.}
To evaluate the spoofing accuracy, we consider victim radars with configurations C and D from Table~\ref{table_simulation_spoofing_victim_configurations}, representing common real-world configurations. As in Section~\ref{Victim_Parameter_Sensing}, we set the attacker's relative position uniformly at random from the intervals [\m{20}, \m{100}]
%\todo{check}
and velocity [\mPers{-10}, \mPers{10}]
%\todo{check}
to evaluate performance across varying positions and velocities. 
%\tingjun{again, what is the distribution? what does random mean?}
%\david{addressed}

We evaluated the spoofing accuracy for 100 different desired spoofing ranges ($\RangeSpoof$) and velocities ($\VelocitySpoof$), 
%. The spoofing ranges ($\RangeSpoof$) were 
uniformly selected at random from the intervals [\m{50}, \m{100}] %while the spoofing velocities ($\VelocitySpoof$) were uniformly selected at random from the interval 
and [\mPers{$-$25}, \mPers{25}], respectively; the test case distributions are summarized in Appendix~\ref{Appendix_Simulation_Details_Spoofing_Tests}. For each trial, we simulated a total number of 10 radar frames, with the attack starting on the 6th frame. As the first 5 frames were used to sense the victim's parameters, 
% \tingjun{why 6-th? does this mean the first 5 frames are used for sensing?}
% \david{Addressed}
each trial featured 5 attack frames.
%\todo{check} 
 %We also verify that the spoofed objects exhibited realistic propagation. %Fig.~\ref{fig_simulated_false_positive_attack_progression} presents an example attack progression for an FP attack.
%tingjun{why are we referring to fig 5? there is also no timeline in fig 5}
%\david{Fig 5 now points to FP attack proression}

%\tingjun{is this the same way of simulating the attack as you did in the pervious section?}
%\david{addressed at the start of this section}
\begin{figure}[!t]
     \centering
     \begin{subfigure}[b]{0.48\columnwidth}
         \centering
         \includegraphics[width=\columnwidth]{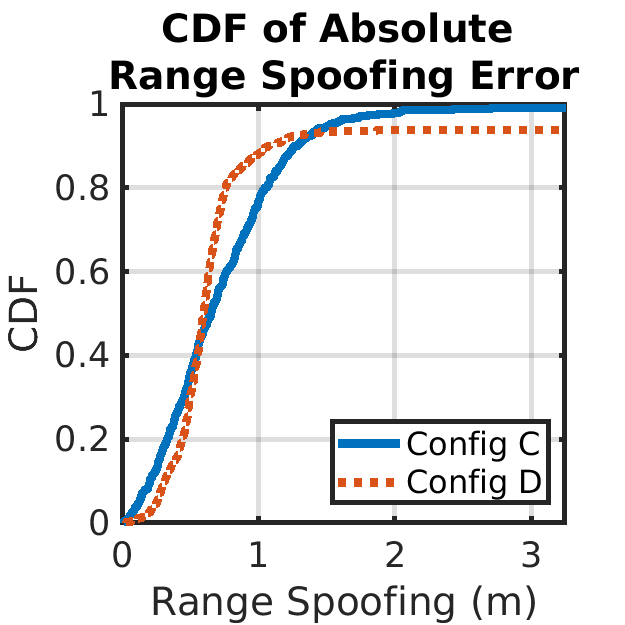}
     \end{subfigure}
     \hfill
     \begin{subfigure}[b]{0.48\columnwidth}
         \centering
        \includegraphics[width=\columnwidth]{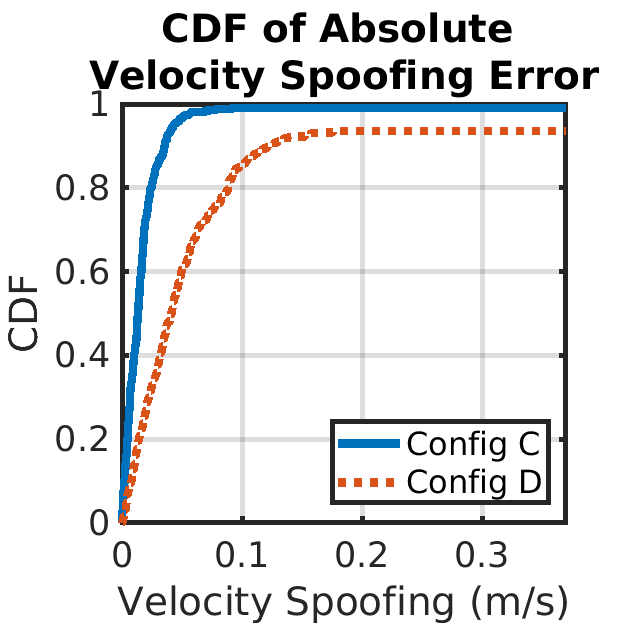}
     \end{subfigure}
     % \vspace{-4pt}
    \caption{CDFs of absolute error for spoofing accuracy.}    \label{fig_simulation_spoofing_error_cdf}
\end{figure}
\begin{table}[!t]
  \caption{Absolute error of attacker spoofing accuracy.}\label{table_simulation_attacker_spoofing_accuracy}
  % \vspace{-6pt}
  \centering
  \begin{tabular}{cccc}
    \toprule
    Configuration & Metric & Mean Absolute Error & 90th Percentile\\
    \midrule
    C & Range & 1.49 m & 1.28 m \\
     & Velocity & 0.15 m/s & 0.04 m/s\\
    \midrule
    D & Range & 4.29 m & 1.09 m \\
     & Velocity & 1.5 m/s & 0.12 m/s\\
  \bottomrule
\end{tabular}
\end{table}

\vspace{1.0ex}
\noindent\textbf{Results.}
Out of the 500 testing frames in the 100 different scenarios, \textbf{\emph{over 90\% of the frames resulted in successful, highly accurate attacks}}. 
%\tingjun{in which cases this would fail in simulations?}
%\david{I tried to better explain in the following sentence. This failed in the simulation when the parameter was insufficiently accurate. Most of this explanation builds off of the previous section, so let me know if I need to add more details to explain inaccuracies in parameter estimation}
%\tingjun{talk about good results first}
%\david{Addressed}
%\tingjun{again, it is not clear to me why there would be imperfections in simulations, unless we manually added imperfections or use some random noise model for signal propagation, etc. We need to be more clear}.
%\david{In the previous section, I added further clarification on the random noise model for signal propagation used, and I tried to better clarify why imperfections existed in the simulation as well. Let me know if there are still more needed here, and I'll fix it.}
Of the successful attacks, Fig.~\ref{fig_simulation_spoofing_error_cdf} shows the CDFs for the absolute range and velocity spoofing errors, and the statistics are summarized in Table~\ref{table_simulation_attacker_spoofing_accuracy}. In particular, 90\% of the successful attacks had the spoofed range within \m{1.28} 
%\tingjun{fixed a number of digits after the decimal point and be consistent with it, I think x.xx is good enough}
%\david{I'll fix the other tables to be to 2 decimal points}
of the desired range ($\RangeSpoof$) and the spoofed velocity within \mPers{0.12} of the desired velocity ($\VelocitySpoof$); %note that 
the absolute velocity spoofing error is significantly lower than the absolute range spoofing error since the velocity spoofing does not depend on the estimated victim chirp period or the predicted frame start time. Also, the mean absolute error for Config~D was relatively high as less than 5\% of trials have spoofing errors significantly larger than the rest of the trials. Finally, the remaining inaccurate spoofing attacks resulted from insufficiently accurate victim parameter estimations as discussed in Section~\ref{Victim_Parameter_Sensing}.
%\todo{check, removed the last sentence as it was the repeat of the `bold' one} 
%Overall, the results show that our spoofing attacks are successful and highly accurate over 90\% of the time.

%%%%%
%%%%%
\subsection{Attack Effectiveness Assessment}

\noindent\textbf{Setup.}
% We also evaluate the attack effects on the victim's probability of detection (PD)\footnote{Probability of detection = 1 - the probability of false negative event.} and probability of false alarm (PFA).\footnote{Probability of false alarm is the probability of a false positive occurring.} 
We also evaluated the attack effect on the victim's PD\footnote{Probability of detection = 1 - the probability of false negative event.} and PFA\footnote{Probability of false alarm is the probability of a false positive occurring.}, the traditional metrics for assessing radar detection performance.
For each configuration in Table~\ref{table_simulation_spoofing_victim_configurations}, we performed 400 different simulations for the FP attack, the FN attack, and the base case without attack. 
Previously, we demonstrated that our framework accurately estimated a victim's parameters, inserting spoofed signals regardless of the relative position and velocity of the attacker and victim. Now, we show that our attacks are successful regardless of the relative position and velocity of a target-of-interest in the environment. 

To maintain a consistent starting point for each simulation, we simulated the attacker \m{75} away from the victim with a relative velocity of \mPers{2}. When evaluating the effectiveness of our FP attacks on a victim's PFA, we selected the desired spoofing range and velocity uniformly at random from the intervals [\m{50}, \m{100}] and [\mPers{$-$25}, \mPers{25}], respectively.
%\todo{check -- removed the repeated last sentence} 
% Finally, the simulated environment was the same as the one described in Section~\ref{Victim_Parameter_Sensing}. 
%\tingjun{why the two ranges are the same? you add an object at the location of the attacker?}
%\davidToDo{Addressed. Repeating the FP simulations right now to address this}

For each trial, we simulated an existing target with velocity ($\VelocityTarget$) uniformly chosen at random from the interval [\mPers{$-$35}, \mPers{35}] (\mPers{35} is roughly 78~mph), and a starting point ($\RangeTarget$) uniformly selected from the interval [\m{5}, \m{143}].
The radar cross-section of the target in each trial was set using the same method described in Section~\ref{sec:sim_env}. 
%\tingjun{you had all kinds of "randomly chosen" but this doesn't say anything, you have to specify the distribution, etc.}
%\david{Fixed to clarify the distribution}
For all the trials, we recorded a FN outcome if there was a real target at a specified location but the victim radar did not detect anything within 3$\RangeRes$ and 3$\VelocityRes$ of an target location. 
%\tingjun{is this $3 \RangeRes$? I think you need to use notations etc more often, otherwise the concept is easily lost}
%\david{Addressed}
We record a FP outcome if the radar detects another object that is not located within 3$\RangeRes$  or 3$\VelocityRes$ of the actual target. 
%\tingjun{why before it's only range bin but now there is range or velocity bin?}
%\david{Fixed}
Thus, it is possible for both a FP and an FN event to occur in the same trial.
%\todo{check} 

\begin{figure}[!t]
     \centering
     \begin{subfigure}[b]{0.48\columnwidth}
         \centering         \includegraphics[width=\textwidth]{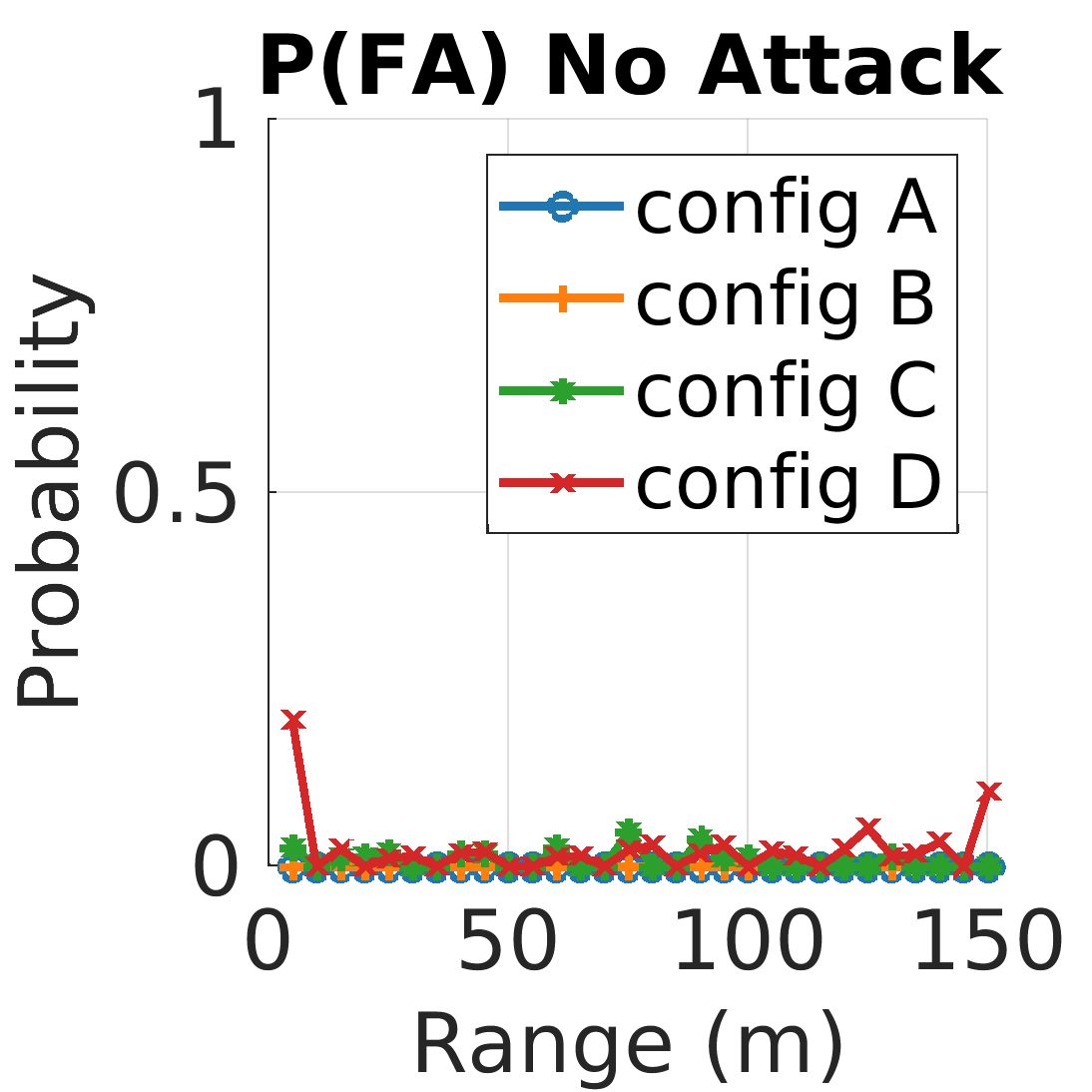}
         % \vspace{-4pt}
         \caption{No Attack}
         \label{fig_simulation_p_fa_no_attack}
     \end{subfigure}
     \hfill
     \begin{subfigure}[b]{0.48\columnwidth}
         \centering
  \includegraphics[width=\textwidth]{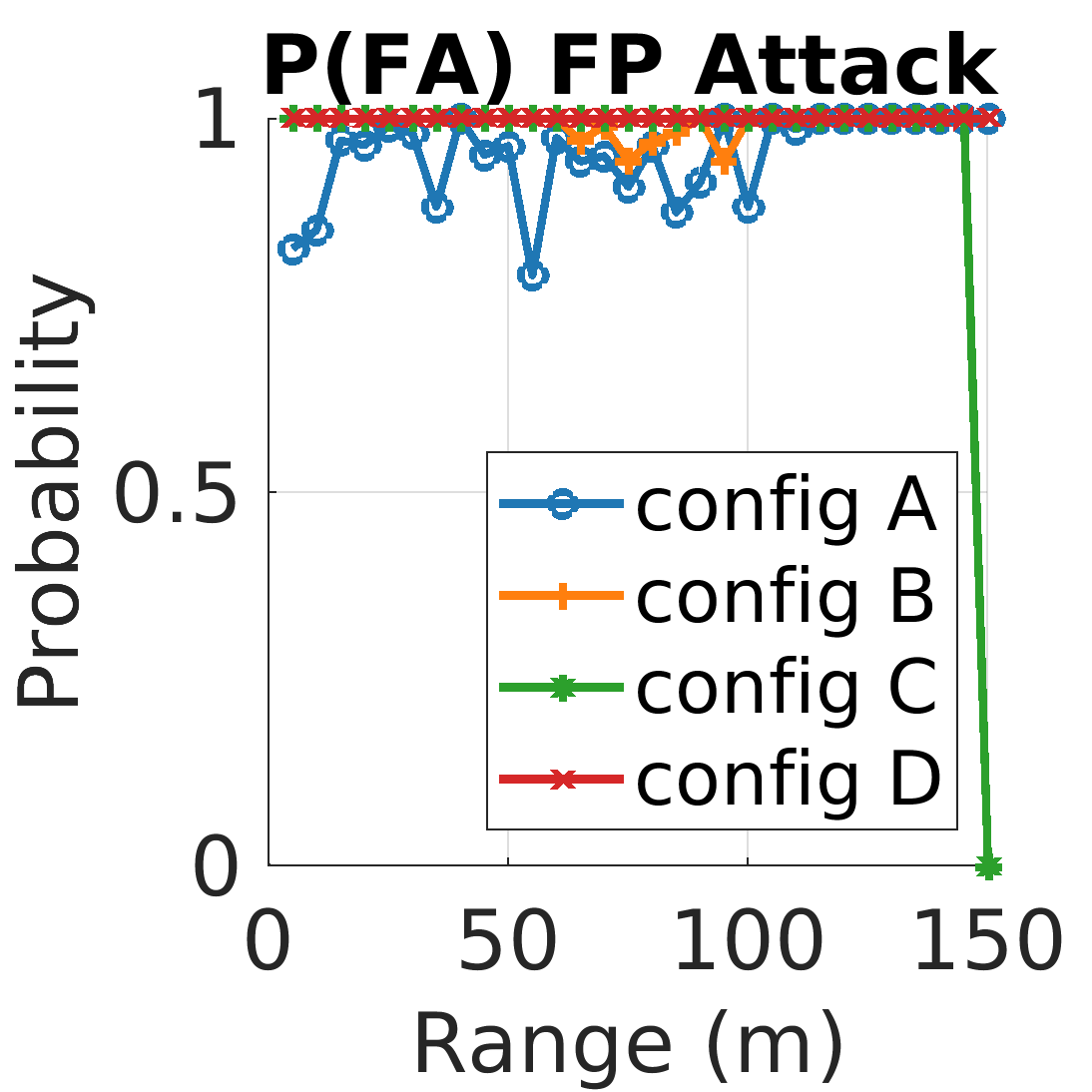}
        % \vspace{-4pt}
         \caption{Spoof. FP~Attack}
         \label{fig_simulation_p_fa_spoof}
     \end{subfigure}
     % \vspace{-2pt}
     \caption{Attack effectiveness on probability of false alarm.}\label{fig_simulated_attack_effectiveness_p_fa}
    % \vspace{-8pt}
\end{figure}

\begin{figure}[!t]
     \centering
      \begin{subfigure}[b]{0.48\columnwidth}
         \centering
         \includegraphics[width=\textwidth]{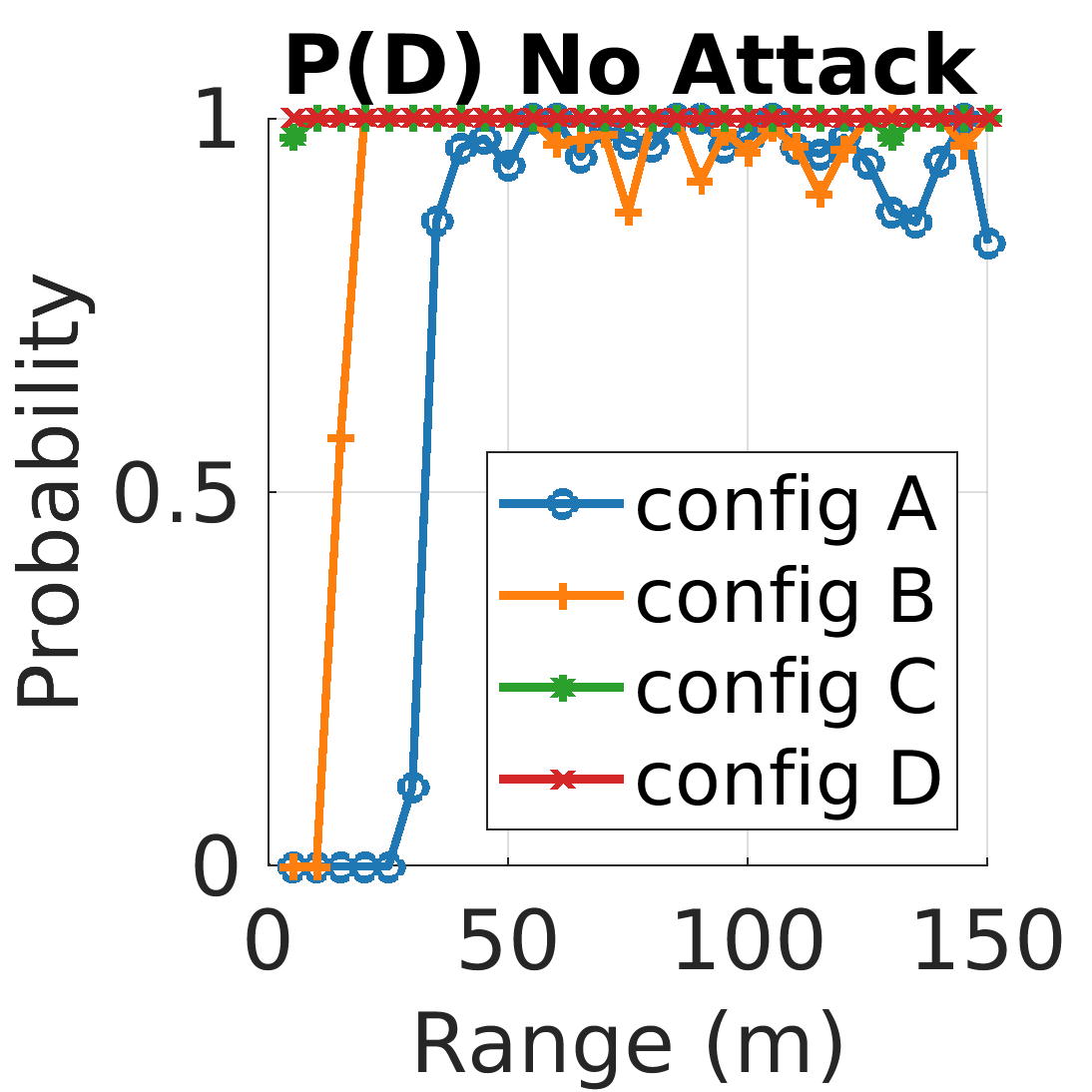}
         % \vspace{-14pt}
         \caption{No Attack}
         \label{fig_simulation_p_d_no_attack}
     \end{subfigure}
     \hfill
     \begin{subfigure}[b]{0.48\columnwidth}
         \centering
\includegraphics[width=\textwidth]{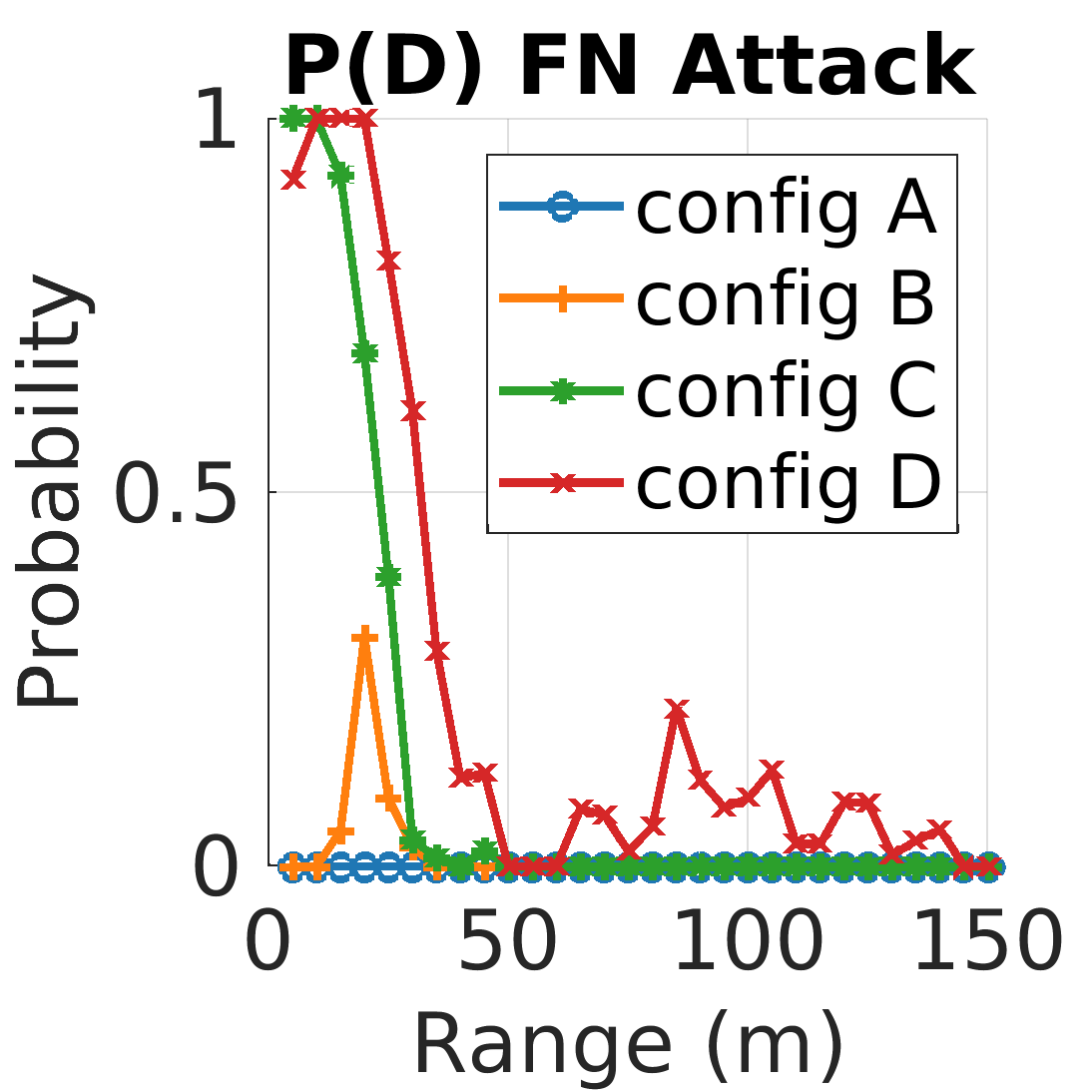}
         % \vspace{-14pt}
         \caption{FN Attack}
         \label{fig_p_d_FN}
     \end{subfigure}
     % \vspace{-2pt}
        \caption{Attack effectiveness on probability of detection.}
        \label{fig_simulated_attack_effectiveness_p_d}
\end{figure}

\vspace{1.0ex}
\noindent\textbf{Results.}
Figs.~\ref{fig_simulated_attack_effectiveness_p_fa}  and~\ref{fig_simulated_attack_effectiveness_p_d} %report the results for evaluating 
show the obtained attack effectiveness; the ``Range'' axis corresponds to the range of the existing target.
%\tingjun{why can't the axis just be range? if its range BIN shouldn't it be unitless??}
%\davidToDo{Update plot axis}
To estimate the PD and PFA, we grouped trials into 1 of 30, \m{5} range bins, i.e., the first range bin contains the results corresponding to a target within the \m{0--5}~range.
%\todo{Check}


\paragraph{FP Spoofing Attacks}
%\david{Does this assessment/section make sense or would I instead be better off just focusing on the FN attack results here?}
Fig.~\ref{fig_simulated_attack_effectiveness_p_fa}(a) shows the PFAs for each victim radar configuration without attacks -- all have very low PFAs ($<5\%$) in this case. Fig.~\ref{fig_simulated_attack_effectiveness_p_fa}(b) shows the PFAs %when the FP spoofing attack is performed 
under the FP spoofing attacks -- the PFAs significantly increased, %Digging deeper, this shows that 
and
the attacker was capable of adding a spoofed (i.e., fake) object into the victim radar's point cloud regardless of the location of the existing (i.e., real) object.~This, combined with the results from Fig.~\ref{fig_simulation_spoofing_error_cdf}, shows that FP attacks~can successfully insert fake objects very close to the desired locations, no matter the positions of other vehicles in the scene. % demonstrates that FP attacks can accurately and successfully spoof a target at all locations.
%\todo{check}

\paragraph{FN Attacks} 
Fig.~\ref{fig_simulated_attack_effectiveness_p_d}(a) shows the PD for each radar configuration when no attack is present -- configurations A and B are unable to detect targets at close ranges due to their poor range resolution ($\RangeRes$) and comparatively high CFAR minimum detection range ($\RangeMin$)  
%the fact that the CFAR detection requires a guard region around each cell under test to estimate the relative noise floor~\cite{katzlberger_object_2018,rohling_radar_1983,rohling_ordered_2011}.
(Table~\ref{table_simulated_victim_cfar_detection_regions} in Appendix~\ref{Appendix_Simulation_CFAR_Detection_Region}). 
%\tingjun{why is PD a function of the range bin?? are you simulating different objects at different ranges and then aggregating all results in one figure? or is it the range bin where the victim is supposed to see only a single object?}
%\david{For the false negative experiments, we simulate objects within each of the range bins (5m to 143 m) and the attacker's goal is to cause the victim to not detect that object. PD is a function of the range bin as PD naturally decreases as a function of range (lower SNR as range increases). Here, we are showing that our attacker dramatically reduces PD compared to nominal performance. I've tried to add a sentence at the end of this paragraph to further clarify why PD is a function of range bin}

Fig. \ref{fig_simulated_attack_effectiveness_p_d}(b) shows the PD for each victim configuration when the FN attack was applied -- the attack significantly decreased each radar's PD, with a steep decline in the victim's PD for targets roughly \m{25} away; the drop at \m{25} occurs because it is roughly the point where the power received from the attacker is equal to the power received from the target reflection. 
While it is expected for PD of a real target to slightly decrease with range (longer ranges experience greater path loss resulting in a reduced SNR), our results show that the FN attacks significantly impact PD compared to operation without an attack. Overall, the results demonstrate that we consistently caused FN events in the victim's radar.%\todo{check}
% employing CA-CFAR detectors. 

%% file: 07_Physical_Implementation.tex
\section{SDR-Based Physical Implementation}
\label{Physical_Implementation}

We now introduce a real-time prototype implementations of {\name} and a victim radar using SDR platforms. Additionally, we validate our prototype's performance on 600 real-world experiments.  Section~\ref{Real_World_Case_Studies} then presents results from multiple real-world case studies.

%%%%%
%%%%%
\subsection{Implementation on an SDR Platform}
\label{Physical_Implementation_Experimental_Hardware}

\begin{figure}[!t]
    \centering
    \includegraphics[width=0.998\columnwidth]{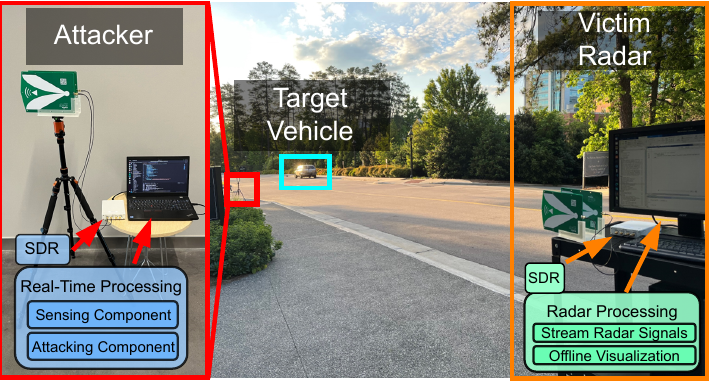}
    % \vspace{-2pt}
    \caption{Physical prototype and setup for some case studies.}
    \label{fig_case_studies_stationary_attack_scene}
\end{figure}

We developed a victim radar and a {\name} prototype using USRP B210 SDRs, which are controlled by host laptops via the C++-based USRP Hardware Driver (UHD)~\cite{ettus_research_usrp_nodate}, as illustrated in Fig.~\ref{fig_case_studies_stationary_attack_scene}; the {\name} prototype alone required 4,500 lines of code. % to implement the \name prototype alone. 
Due to the limitations of available hardware %of the USRP B210, computer processing power, and RF antennas
(e.g., the frequency range of [\MHz{70}, \GHz{6}] and maximum sampling rate of \MHz{56} for the USRP B210),
%\todo{check}
we consider an operating frequency ($\ChirpFreqStart$) of \GHz{1.5} and a sampling rate of \MHz{25}; this corresponds to a victim range resolution ($\RangeRes$) of $\sim$\m{6} and the maximum timing accuracy of the \name~attack framework of $\sim$\nsec{40}. We apply longer chirps and frames to achieve a realistic velocity resolution of $\sim$\mPers{0.8}.
% Additionally, the operating frequency of {1.5}\thinspace{GHz} meant we had to use longer chirps and frames to achieve a  more realistic velocity resolution of $\sim${0.8}\thinspace{m/s}. Further impacting the velocity spoofing capability of our prototype, \cite{ettus_research_b200b210b200minib205mini_nodate} notes that the USRP B210 has a phase noise of 1.0 degrees RMS at {3.5}\thinspace{GHz}. Using \eqref{eq_velocity_computation}, this corresponds to $\sim${0.5}\thinspace{m/s} of potential error in the spoofing velocity due to phase noise. 
While our prototype implementation is constrained by the hardware limitations, it can easily be extended to a full-scale implementation with the use of more capable (and expensive due to high-frequency SDR) hardware platforms.
%\todo{check} 

For our experiments, the victim radar transmits a series of %pre-computed 
FMCW chirps and records the received signal (i.e., the reflected chirps), which is then processed offline to obtain the Range-Doppler response and detect objects using the pipeline from Fig.~\ref{fig_FMCW_pipeline}.
%\tingjun{say following pipeline described in Section/Fig XXX}
%\david{Addressed}
%\tingjun{why an example? is this everything that we did?}
%\david{removed as its irrelevant for this section}
The \name~prototype, \emph{in real-time}, estimates the victim radar's chirp slope ($\EstimatedChirpSlope$), chirp period ($\EstimatedChirpDuration$), and frame duration ($\EstimatedFrameDuration$) %using the method 
as described in Section~\ref{Attack_Design}. 
%\tingjun{using pipeline/method described in Section/Fig XXX, bring back the notation}
%\david{Added citation}
Based on the estimated parameters, the attacker designs and transmits the corresponding signal for launching a FP, FN, or translation attack using \eqref{eq_FP_attack_chirp} and \eqref{eq_FN_attack_chirp}.
%\tingjun{refer to Section/Equation/Fig. XXX as fit}
%\david{Addressed to refer to equations}
%We highlight that the 
The {\name} prototype is the \textbf{\emph{first to demonstrate the feasibility of launching real-time black-box FP, FN, and translation attacks on a real-world system}}. 

% \paragraph{Victim Implementation} 
% Fig~\ref{fig_case_studies_stationary_attack_scene} presents an overview of the prototype victim and attacker we implemented using the USRP B210 SDR platform. For our prototype, the victim transmitted a series of pre-computed FMCW chirps for each radar frame and saved the received signals to a file; we then used an offline MATLAB script to visualize the Range-Doppler response and the detected objects. Table~\ref{table_experimental_configuration} features an example of a victim configuration we utilized in our experiments.
% \vspace{-2pt}
% \paragraph{Attack Implementation}
%  Our prototype attack implementation on the USRP B210 estimated the victim's chirp period, frame rate, and chirp slope in real-time without any offline processing; these estimates were then used to compute and transmit attacking signals for the attacks introduced in Sec.~\ref{Attack_Design} in real-time. Thus, our prototype is the \emph{first to demonstrate the feasibility of launching simultaneous black-box FP, FN, and translation attacks in real-time on a real-world system.} 

\subsection{Physical Evaluation of Parameter Estimation}
\label{sec:physical_sensing_verification}

%As in Sec.~\ref{Victim_Parameter_Sensing}, we performed a rigorous evaluation of our prototype's parameter estimation capability to validate that its estimates were accurate enough to enable effective attacks. \tingjun{opening/intro paragraph like this can be removed once the structure of the subsectino is clear, see what I did below as an example.}

\noindent\textbf{Setup.}
We utilized 500 different victim configurations to validate the {\name} prototype's accuracy of estimating the victim radar's parameter. Due to the hardware limitations, we considered victim configurations with a chirp bandwidth of up to \MHz{25}. The victim chirp slope and duration ($\ChirpSlope$ and $\ChirpDuration$) were chosen uniformly at random from the intervals [\MHzPerus{0.05}, \MHzPerus{0.53}] and [\usec{50}, \usec{500}], respectively (details in Appendix~\ref{Appendix_Experimental_Details}). The maximum chirp slope was small due to the maximum chirp bandwidth of \MHz{25}. For each victim configuration, \name's sensing module estimates the victim radar's chirp slope ($\EstimatedChirpSlope$), chirp period ($\EstimatedChirpDuration$), and frame duration ($\EstimatedFrameDuration$) over a series of 7 frames. Also, the prototype initiated attacks on the 7-th frame, continuing %sensing performance 
parameter estimation while performing real-time~attacks.%\todo{check}
% As in Sec.~ \ref{Victim_Parameter_Sensing}, we recorded the estimated chirp periods and chirp slopes after the 7th frame.

%\subsubsection{Results}

\begin{figure}[!t]
    \centering
    \includegraphics[width=0.94\columnwidth]{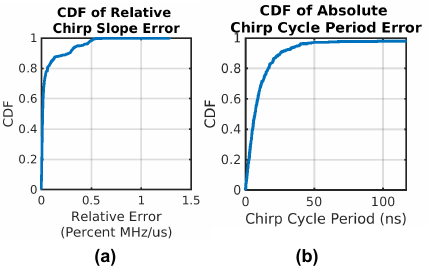}
    % \vspace{-8pt}
    \caption{CDFs of the absolute errors for the parameter estimation on SDR-based physical platforms.}
    \label{fig_experimental_parameter_estimation_cdf}
\end{figure}
\begin{table}[!t]
  \caption{Key parameter estimation error metrics for the implemented physical prototype.}
  \label{table_experimental_parameter_estimation_results}
  % \vspace{-6pt}
  \centering
    \begin{tabular}{cccl}
        \toprule
        Parameter & Error Type & Mean Error & $95^{th}$ Percentile\\
        \midrule
        Chirp Period & Absolute & \nsec{18.95} & \nsec{39.09}\\
        Chirp Slope & Absolute & \MHzPerus{0.0025} & \MHzPerus{0.00175} \\
         Chirp Slope & Relative & $0.0658 \%$ & $0.403 \%$ \\
        \bottomrule
    \end{tabular}
    % \vspace{-2pt}
\end{table}
%
%Fig.~\ref{fig_experimental_parameter_estimation_cdf} presents the obtained CDFs of the relative errors for the chirp slope and the absolute errors for chirp period estimations, whereas Table~\ref{table_experimental_parameter_estimation_results} summarizes the key statics from the distribution. \tingjun{again, once you have the strcture of the subsection we don't need such a paragraph or subsubsection. this is repeated below anyway.}

\vspace{1.0ex}
\noindent\textbf{Chirp period estimation.}
Table~\ref{table_experimental_parameter_estimation_results} and Fig.~\ref{fig_experimental_parameter_estimation_cdf}(a) summarize the results for the chirp period estimation over all physical trials -- in 95\% of trials, the estimate of the chirp period ($\EstimatedChirpDuration$) is within \nsec{39.09} of the actual chirp period ($\ChirpDuration$). Compared with the full-scale simulation-based results from Section~\ref{Victim_Parameter_Sensing} (95\% of trials had an absolute error less than \nsec{0.59}), the timing accuracy decrease by roughly two orders of magnitude is attributed to the described hardware constraints of the prototype system. However, these results indicate that 95\% of our chirp period estimates are within one sampling period of the actual victim radar's chirp period.
%\todo{check}
% Thus, we demonstrate that our parameter estimation subsystem is realizable. We will show that these estimates are sufficiently accurate to launch successful attacks later in this section.

\vspace{1.0ex}
\noindent\textbf{Chirp slope estimation.}
Table~\ref{table_experimental_parameter_estimation_results} and Fig.~\ref{fig_experimental_parameter_estimation_cdf}(b) summarize the obtained results of physical evaluation -- over 95\% of the trials result in estimated chirp slope values ($\EstimatedChirpSlope$) that are within \MHzPerus{0.0017} (relative error of 0.403\%) of their actual values ($\ChirpSlope$). This absolute value is quite low in part because all of the tested radar victim configurations had relatively small slopes. However, the relative value indicates that the accuracy was similar to what was observed in our simulation results (Section~\ref{Victim_Parameter_Sensing}).
%\todo{check}
Again, we attribute the order of magnitude increase in relative slope estimation error to the lower sampling bandwidth of \MHz{25}, impacting the maximum achievable resolution when generating a spectrogram of the received signal chirps.
% Still, these results and the results in the following sections show that the developed physical prototype estimation subsystem is capable of accurately estimating a victim's parameters in over 95\% of cases.

%%%%%
%%%%%
\subsection{Physical Evaluation of Spoofing Accuracy}
\label{sec:physical_spoofing_verification}

\begin{table}[!t]
\begin{center}
    \caption{Victim radar configuration used for evaluation of the physical prototype and case studies 
    %\tingjun{something is weird, the bottom line is cut half way through}
    }
    \label{table_experimental_configuration}
    % \vspace{-6pt}
    \begin{tabular}{ 
    m{.12\columnwidth}
    m{.12\columnwidth}|
    m{.40\columnwidth}}
    \toprule
    Parameter & (Units) & Experimental Configuration \\
    \midrule
    $\ChirpFreqStart$ & GHz & 1.5 \\
    $\ChirpBW$ & MHz & 25.00 \\
    $\ChirpSlope$ & MHz/us & 0.05 \\
    $\ChirpDuration$ & us & 501.12 \\
    $\NumChirpsPerFrame$ & & 256 \\
    $\RangeRes$ & m & 6.09 \\
    $\RangeMax$ & m & 1,558.92 \\
    $\RangeMin$ & m & 24.35 \\
    $\VelocityRes$ & m/s & 0.78 \\
    $\VelocityMax$ & m/s & 99.71 \\
    \bottomrule
    \end{tabular}
\end{center}
\end{table}

\noindent\textbf{Setup.}
We evaluate spoofing accuracy over 100 unique trials where each real-time trial involved 15 attacking frames. Table~\ref{table_experimental_configuration} summarizes the victim configuration used for all trials; this configuration used a longer chirp duration ($\ChirpDuration$) to achieve more realistic velocity resolution ($\VelocityRes$) given the lower operating frequency ($\ChirpFreqStart$) of the physical prototype. The spoofing range ($\RangeSpoof$) and velocity ($\VelocitySpoof$) for each trial was uniformly chosen at random from the interval [\m{60}, \m{200}] and [\mPers{-25}, \mPers{25}] (see Appendix~\ref{Appendix_Experimental_Details} for distribution). 
%Finally, each trial was performed in real-time with only offline processing being used to determine the location where the victim perceived each spoofed object. 

%
\begin{table}[!t]
  \caption{Absolute error of the attack spoofing accuracy.}
  \label{table_experimental_physical_spoofing_accuracy}
  % \vspace{-6pt}
  \centering
  \begin{tabular}{ccc}
    \toprule
    Metric & Mean Absolute Error & 90th Percentile\\
    \midrule
    Range & 7.53 m & 9.67 m \\
    Velocity & 1.42 m/s & 1.80 m/s\\
  \bottomrule
\end{tabular}
\end{table}

\begin{figure}[!t]
     \centering
     \begin{subfigure}[b]{0.486\columnwidth}
         \centering
         \includegraphics[width=\columnwidth]{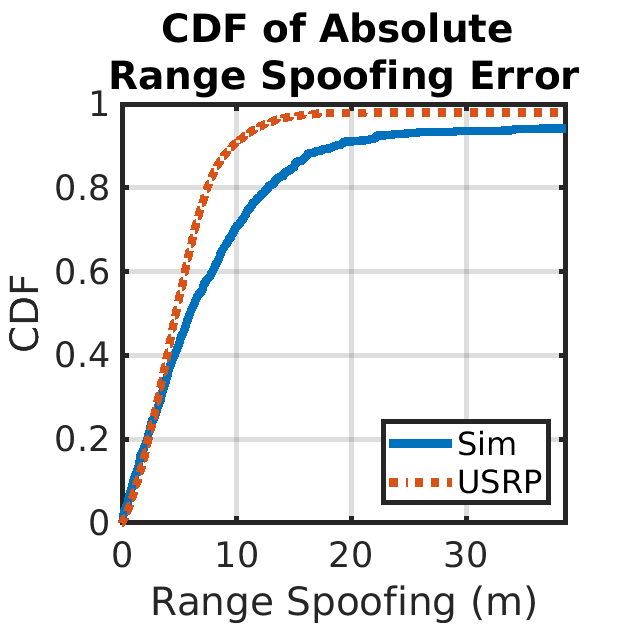}
     \end{subfigure}
     \hfill
     \begin{subfigure}[b]{0.486\columnwidth}
         \centering
         \includegraphics[width=\columnwidth]{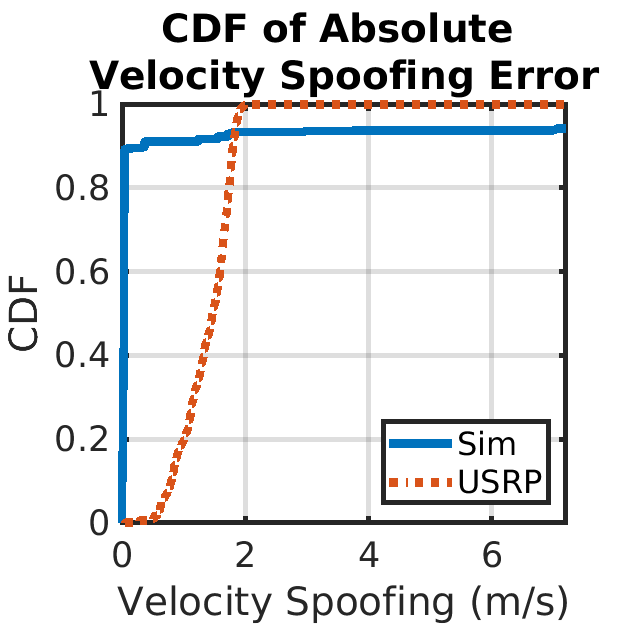}
     \end{subfigure}
        \caption{CDFs of absolute error of spoofing accuracy obtained in physical experiments (orange) and corresponding simulation (blue).}
        \label{fig_experimental_spoofing_error_cdf}
\end{figure}

\vspace{1.0ex}
\noindent\textbf{Results.} Fig.~\ref{fig_experimental_spoofing_error_cdf} reports the obtained CDFs for the absolute range and velocity errors, whereas Table~\ref{table_experimental_physical_spoofing_accuracy} summarizes the relevant statistics -- 90\% of trials had %a detected 
the obtained range within \m{9.67} of the desired spoofing range ($\RangeSpoof$) and %a detected 
the obtained velocity \mPers{1.80} of the desired spoofing velocity ($\VelocitySpoof$).
%\todo{check} 
To compare the experimental results with the simulation-based results presented in Section~\ref{Large_Scale_Evaluation}, we simulated the exact same set of physical scenarios using the simulated environment from Section~\ref{Victim_Parameter_Sensing}. In Fig.~\ref{fig_experimental_spoofing_error_cdf}, the simulated results appear in blue while the  results obtained in physical experiments appear in orange. We highlight how our prototype spoofed an object's range slightly more accurately than our simulation predicted. While we observe that the prototype's velocity spoofing was less accurate than the simulations predicted, we attribute this discrepancy to the phase noise in the USRP B210,\footnote{The USRP B210 has a phase noise of 1.0 degrees RMS at {3.5}\thinspace{GHz}~\cite{ettus_research_b200b210b200minib205mini_nodate},
corresponding to $\sim${0.5}\thinspace{m/s} of potential error in the spoofing velocity due to phase noise; this follows from~\eqref{eq_velocity_computation}.} which our simulations do not account for. 

In summary, the results from our real-world physical experiments demonstrate that {\name}
%can feasibly be implemented on a real-time system. 
 %Additionally, our prototype
estimates a victim's parameters and inserts spoofed objects with the anticipated level of accuracy given the hardware limitations.
%\todo{check this+previous paragraph}

%% file: 07a_Real_World_Case_Studies.tex
\section{Real World Case Studies}
\label{Real_World_Case_Studies}

\begin{figure*}[!t]
    \centering
    \includegraphics[width=0.996\textwidth]{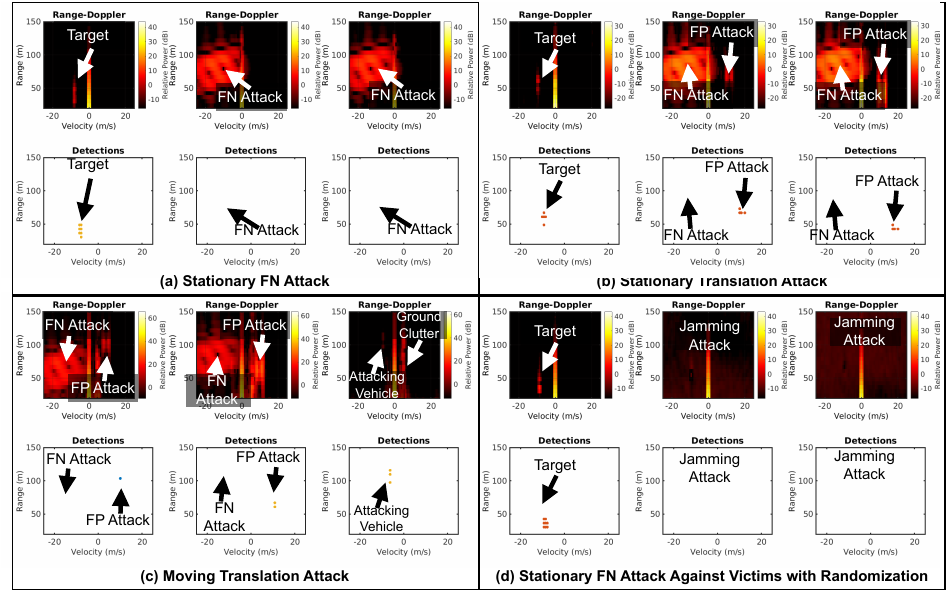}
    % \vspace{-4pt}
    \caption{Attack progressions for real world case studies.}
    % \vspace{-6pt}
    \label{fig_case_studies_combined_attack_timelines}
    % \vspace{-8pt}
\end{figure*}

We now demonstrate the \emph{real-world} capability of {\name} through several real-world case studies. 
We start by demonstrating FN and translation attacks against stationary victims followed by a demonstration of a translation attack on a moving victim. 
%Finally, we demonstrate how our proposed framework is still effective against victims that employ basic parameter randomization. 
To the best of our knowledge, 
%no previous works have performed these case studies. Thus, 
\emph{\textbf{this it the first work to demonstrate each of the following attack capabilities in realistic case studies}}. Results and details from these and additional case studies, including time-synchronized videos, are available on the project website \cite{Project_Website}.%\todo{check}
\begin{figure}[!t]
    \centering
    \includegraphics[width=0.998\columnwidth]{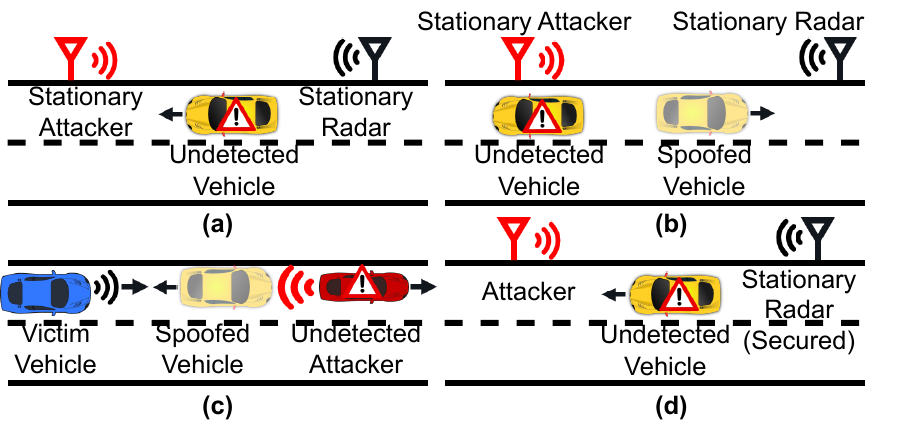}
    % \vspace{-5pt}
    \caption{Considered case study threat scenarios for (a)~stationary FN attack, (b)~stationary translation attack, (c)~moving translation attack, and (d)~stationary 'Jamming' attack.}% (graphics from \cite{blue_car,red_car,yellow_car,wireless_signal,danger_sign})}
    \label{fig_case_studies_threat_models}
\end{figure}

\subsection{Stationary Case Studies}
We first present case studies where a stationary attacker is set up to attack a stationary victim %. Here, the victim attempts 
trying to detect objects on the road. Meanwhile, the attacker estimates the victim radar's parameters in real-time and then simultaneously launches the {\name} attacks. % introduced in Section ~\ref{Attack_Design}. 
This experimental setup is commonly found in the real world including infrastructure sensors detecting vehicles at stoplights and stopped vehicles sensing oncoming traffic prior to pulling out of a parking lot.
%Thus we demonstrate , accurately detecting one's surroundings is critical to safety on the roads.

\vspace{1.0ex}
\noindent\textbf{Setup.}
Fig.\ref{fig_case_studies_stationary_attack_scene} illustrates the experimental setup while Fig.~\ref{fig_case_studies_threat_models}(a) and Fig.~\ref{fig_case_studies_threat_models}(b) portray the threat scenarios used for the stationary case studies. Here, the attacker and victim were placed \m{15} apart from each other. A real target vehicle then drove away from the victim at approximately \mPers{9} ($\sim$20 mph). Finally, the victim radar employed the configuration described in Table~\ref{table_experimental_configuration}.

\begin{figure}[!t]
    \centering
    % \vspace{-4pt}
    \includegraphics[width=0.998\columnwidth]{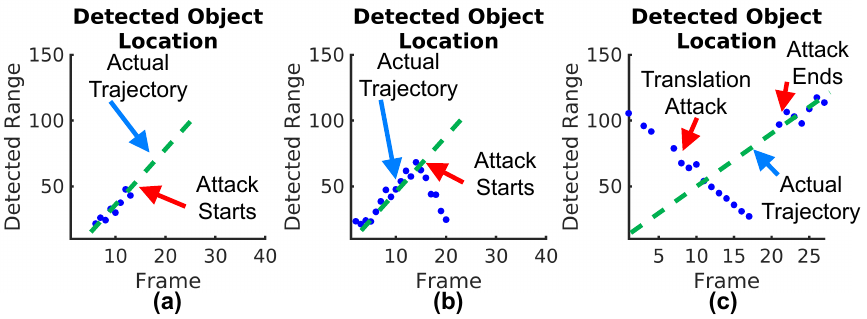}
    % \vspace{-4pt}
    \caption{Radar detections over time for: (a) stationary FN attack; (b) stationary translation attack; and (c) moving translation attack.}
    % \vspace{-6pt}
    \label{fig_case_studies_attack_detections}
\end{figure}

\vspace{1.0ex}
\noindent\textbf{FN attacks.}
% Here, the attacker 
We launched FN attacks with $\RangeSpoof$ and $\VelocitySpoof$ set to \m{75} and \mPers{-5}, respectively. The attack started on the $\textrm{11}^{\textrm{th}}$ sensed frame; this coincided with causing a FN event when the target vehicle was at \m{50} distance. The attack progression is shown in Fig.~\ref{fig_case_studies_combined_attack_timelines}(a) while Fig.~\ref{fig_case_studies_attack_detections}(a) shows the detected target location for each victim radar frame. The first column of Fig.~\ref{fig_case_studies_combined_attack_timelines}(a)
%\todo{was b, but should be a, right?}
presents the victim's perception prior to the attack while the second and third columns present the victim's perception  while under attack. The victim radar's immediately fails to detect the target vehicle once the FN attack is launched. Such a result is incredibly critical as the attack has effectively `removed' an object from the victim's point cloud/scene. 

\vspace{1.0ex}
\noindent\textbf{Translation attack.} Here, we simultaneously launched the FN attack from~\eqref{eq_FN_attack_chirp} and FP attack from~\eqref{eq_FP_attack_chirp}. The FN attack was launched with $\RangeSpoof$ and $\VelocitySpoof$ set to \m{75} and \mPers{-5}, respectively. Simultaneously, the FP attack started at \m{75} while propagating towards the victim with a velocity of \mPers{10}. The attack progression is featured in Fig.\ref{fig_case_studies_combined_attack_timelines}(b) while Fig.~\ref{fig_case_studies_attack_detections}(b) presents the attack detections over time. The first column of Fig.~\ref{fig_case_studies_combined_attack_timelines}(b) presents the victim's perception prior to the attack while the second and third columns present the victim's perception  while under attack. Even though the actual target of interest was \emph{moving away} from the victim for the duration of the experiment, we observe that the victim erroneously perceived that the target vehicle started \emph{moving toward} it once the attack started.
%\todo{check}
Also, note that the power level of the spoofed (fake) object increases as it gets `closer' to the victim radar. Overall, such an attack is incredibly powerful as an attacker can effectively `move' any specific object in the victim radar's point cloud (i.e., perceived scene).

\subsection{Moving Case Studies}
We now demonstrate that {\name} can launch succesful black-box translation attacks from a moving vehicle, which can critically affect safety of autonomous vehicles. % with increased levels of automation/autonomy.
%\todo{check} %. Thus, we show how our attacks can impact safety-critical advanced driver assistance systems including forward collision warning (FCW), blind spot monitoring (BSM), automatic cruise control (ACC), and other autonomous driving applications.

\begin{figure}[!t]
    \centering
    % \vspace{-6pt}
    \includegraphics[width=0.88\columnwidth]{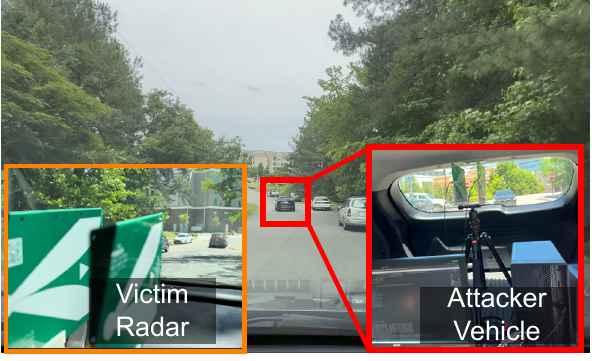}
    % \vspace{-4pt}
    \caption{Experimental setup for the moving vehicles studies.}
    % \vspace{-10pt}
    \label{fig_case_studies_moving_attack_scene}
\end{figure}

\vspace{1.0ex}
\noindent\textbf{Setup.}
Fig.~\ref{fig_case_studies_moving_attack_scene} and Fig.~\ref{fig_case_studies_threat_models}(c) feature the experimental setup and threat scenarios considered in our moving-vehicle case studies. %Here, our prototype attacker 
The {\name} platform was placed in the trunk of the attack vehicle so that it could sense the victim radar's parameters and launch attacks. The victim radar was placed in a separate vehicle that moved independently from the attacker's vehicle.

At the start of the %case study, 
experiment, the attacker and victim began driving forward at \mPers{13} and \mPers{4.5} (30 mph and 10 mph), respectively. Here, the attacker immediately launched the translation attack -- % introduced in Section~\ref{Attack_Design}. 
the FN attack started at a target range of \m{75} and propagated away from the victim radar with the target's velocity of $\sim$\mPers{10}, while the FP attack started at \m{100} and propagated towards the victim radar with velocity of \mPers{12}.

\vspace{1.0ex}
\noindent\textbf{Results.}
The attack progression is featured in Fig.~\ref{fig_case_studies_combined_attack_timelines}(c) while Fig.~\ref{fig_case_studies_attack_detections}(c) reports the detected target locations over time. The first and second columns of Fig.~\ref{fig_case_studies_combined_attack_timelines}(c) presents the victim radar's perception during the translation attack while the third column presents its perception after the translation attack concluded. %We highlight how the attacker launched a successful 
The successfully launched translation attack lead the victim to believe that the attack vehicle was moving \emph{towards} it even though the attack vehicle was actually moving \emph{away} from it. Such an attack is incredibly powerful as the victim failed to detect the attacker's actual location while simultaneously detecting the attacker's fake location; this could lead to very dangerous situations in real-world scenarios. Finally, the victim was only able to detect the actual location of the attacker once the translation attack completes, further demonstrating the effectiveness of the attack.
%\todo{check}

%% file: 08_Discussion_and_future_works.tex
\section{Discussion and Future Work}
\label{Discussion_and_Future_Work}

\subsection{Limitations of {\name}}
\label{Discussion_and_Future_Work_Limitations}

\noindent\textbf{Angle of Arrival (AoA).} 
Modern radars detect an object's range, velocity, and angle of arrival. As dicussed in Section~\ref{Attack_Objectives}, we assume that the attacker is physically located at the desired angle of attack. More versatile attackers would attack a victim from any angle within the victim's field of view. Future works will explore methods for angular spoofing attacks.
%\todo{removed the references to avoid explaining again that they are white box + all the other comments} % such as ~\cite{sun_who_2021}.
%To achieve this, we will explore combining our framework with the white-box angular spoofing attack introduced in~\cite{sun_who_2021}.

\vspace{1.0ex}
\noindent\textbf{CFAR detection.} 
In this work, we focused {\name} attacks on radars employing the  widely-used CA-CFAR detector. However, other CFAR detectors exist, e.g., %the Ordered Statistic CFAR detector (
OS-CFAR~\cite{katzlberger_object_2018,rohling_radar_1983,rohling_ordered_2011}. While we expect that the presented FN and translation attacks can be used for other radar designs, future work will include attack demonstrations against other CFAR detectors.

\vspace{1.0ex}
\noindent\textbf{Physical implementation.} 
As described in Section~\ref{Physical_Implementation}, our physical prototype was limited by the available hardware. A more capable (yet, very expensive) hardware platform could be used to implement a full-scale version. To start, an RF chain capable of converting between baseband frequencies and mmWave frequencies (\GHz{77--81}) could be used, % could be implemented using commercially available parts, 
but %the hardware 
at such high frequencies, with the cost of tens of thousands of dollars. Moreover, generating a spectrogram for the full \GHz{4} bandwidth utilized by automotive radars would require an expensive RFSOC board %capable of achieving 
supporting complex sampling rates of at least {4}\thinspace{Gsps} (such platforms have been used in e.g.,~\cite{peters_arestor_2022}). Future works will seek to develop a full scale system.
%\todo{check}

%%%%%
%%%%%
\subsection{Potential Defenses and Future Works}
\label{Discussion_and_Future_Works_Potential_Defenses}

\noindent\textbf{Parameter randomization.}
Most PHY attacks %pertaining to 
on automotive FMCW radars, including ours, rely on the ability to predict when the victim's next frame will occur. As described in~\cite{amar_fmcw-fmcw_2021,kunert_mosarim_2010,kunert_d15_2010, sun_who_2021}, %one of the best ways to defend 
a defense against such attacks could be introducing small random changes to a radar's chirp period ($\ChirpDuration$), chirp slope ($\ChirpSlope$), and frame duration($\FrameDuration$) at each frame; no commercially available mmWave radars support this. While such a defense could thwart most other spoofing attacks, {\name} framework can be easily modified to launch effective attacks against victims employing such a defense.
%\todo{Check}

Using the same stationary experimental setup from Section~\ref{Real_World_Case_Studies}, we modified our victim to randomize the start time of each frame using a normally distributed offset with a mean of \usec{0} and $3 \sigma$ (3$\times$ standard-deviation) of \usec{3}. We modified the {\name} prototype to detect victims employing parameter randomization
%\todo{removed the footnote, as minor detail; check} %\footnote{We assume timing offsets with a variance of at least $\textrm{\usec{0.1}}^\textrm{2}$.}
using the Liklihood Ratio Test~\cite{kay_estimator-correlator_1998}.
%\todo{also removed the foonote; check} %\footnote{We empirically determined that our sensing component had a timing measurement error variance of $\textrm{\usec{0.015}}^\textrm{2}$.}
Once the attacker detects parameter randomization, it generates an optimal FN attack against randomization (i.e., FN `jamming' attack) leveraging the FN attack from \eqref{eq_FN_attack_chirp} to cover a much larger range ($\sim$\m{1,000}) and velocity ($\sim$\mPers{200}) spread.
%\todo{check the paragraph} 

\begin{figure}[!t]
    \centering
    \includegraphics[width=0.98\columnwidth]{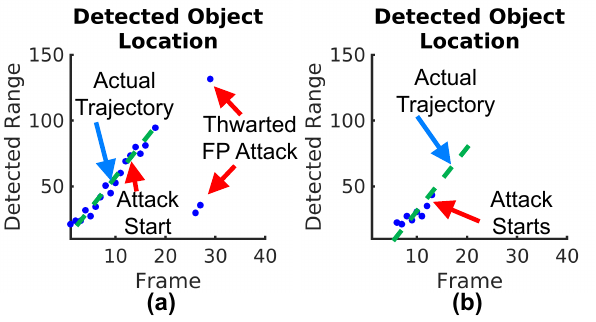}
    % \vspace{-6pt}
    \caption{Detections over time for attacks: (a) without randomization detection; and (b) with randomization detection.}
    \label{fig_case_studies_stationary_jamming}
    % \vspace{-10pt}
\end{figure}

The first column of Fig.~\ref{fig_case_studies_combined_attack_timelines}(d) features the victim's perception when not under attack, whereas the 2nd and 3rd columns present its perception under the optimal FN `jamming' attack. Fig.~\ref{fig_case_studies_stationary_jamming} compares the victim's detections over time when attacked by our standard translation attack and our optimal FN `jamming' attack. 
%Observer how the standard translation attack is thwarted, allowing the victim to still detect the target due to the randomized frame start times. 
We highlight how the optimized FN `jamming' attack still prevents the victim from detecting the target even though standard spoofing attacks may be thwarted in a sense that the accuracy of the FP detections is significantly affected.
%\todo{chekc}  
%While this does not allow us to add spoofed objects or remove specific objects form the victim's point cloud, 
% While not as effective as our primary FP, FN, and translation attacks, such an attack still impacts a victim's ability to detect targets by adding clutter throughout the whole Range-Doppler response.
As an avenue for future work additional analysis of defense mechanisms against {\name} will be performed.
%\todo{CHECK!!} 

% Still, less common methods of parameter randomization could be effective against our proposed framework. For example, the authors of~\cite{moon_bluefmcw_2022} proposed a novel frequency hopping method for FMCW radars that effectively breaks each chirp into several parts and randomly hops between the different parts. This method would thwart all white-box attacks and prevent all black-box attacks from estimating key victim parameters. However, this defense method is complicated to implement and it is not known if it has been deployed on commercial systems. Future works will develop attacks against such advanced defenses.

\vspace{1.0ex}
\noindent\textbf{Multi-sensor fusion.}
While we focus on attacking a single radar sensor, modern vehicles utilize signals from multiple sensors including cameras, LiDARs, and radars. Even if our attack was successful in manipulating a victim radar's data, it is possible that the attack could be thwarted using the victim's other sensors. Still, such a defense is not guaranteed to succeed as works such as~\cite{hallyburton_security_2022} demonstrated successful attacks against victims employing LiDAR-camera sensor fusion. 
%Similar methods could be developed for victims employing radar-camera sensor fusion. 
Future works will investigate PHY radar attacks on vehicles employing radar-camera and radar-LiDAR-camera sensor fusion.
%\todo{check} 

%% file: 09_Conclusion.tex
\section{Conclusion}
\label{09_Conclusion}

In this work, we presented the design of {\name}, a novel black-box physical layer attack framework for mmWave FMCW automotive radars. Unlike previous works that focused solely on `adding' fake objects into a victim radar's point cloud, this is the first work to introduce the false-negative and translation attacks that effectively `\emph{remove}' or `\emph{move}' detections of existing objects in the victim radar's point cloud. Further, all but one of the previous (false-positive only) spoofing attacks assumed prior knowledge of the victim radar's parameters. By comparison, {\name} estimates the victim's chirp period, chirp slope, and frame duration with sufficient accuracy to implement successful attacks over 95\% of the time. We have experimentally validated the feasibility and effectiveness of the proposed attacks by developing a real-time {\name} prototype using SDR platforms. Finally, we have demonstrated the real-world capabilities of {\name}through real-world case studies.

%% file: A0_acknowledgments.tex
\section*{Acknowledgments}
This work is sponsored in part by the ONR under agreements N00014-23-1-2206 and N00014-20-1-2745, AFOSR under award number FA9550-19-1-0169, as well as by the NSF grants CNS-1652544 and CNS-2211944, and the National AI Institute for Edge Computing Leveraging Next Generation Wireless Networks (Athena), grant CNS-2112562.
% Add acknowledgements here

%% file: A1_Appendix_1.tex
\subsection{FMCW Radar Signal Processing Theory} 
\label{Appendix_Preliminaries_on_Radar_Signal_Processing}
\subsubsection{Simplification of the IF Signal} \label{Appendix_Preliminaries_Simplification_of_IF_Signal}
As discussed in \eqref{eq_FMCW_IF_signal_simplified} from Section~\ref{Preliminaries}, the IF signal is obtained by mixing the transmitted signal with the received signal -- i.e., 
\begin{equation} \label{eq_IF_signal_unsimplified}
    \begin{split}
        s_{\textrm{IF}}^{(l)}(t) = & x(t) \cdot y^{*}(t) \\
        = & A_{\textrm{Rx}} \cdot \exp \{ j\left[2 \pi \ChirpFreqStart \cdot t + \pi \ChirpSlope \cdot t^2\right] - \\
        & \hspace{.1cm} j \left[2 \pi \ChirpFreqStart (t-\tDelay) + \pi \ChirpSlope (t - \tDelay)^2  \right] \} + z'(t) \\
        = & A_{\textrm{Rx}} \cdot \exp \{ j [ 2 \pi \ChirpFreqStart \cdot t + \pi \ChirpSlope \cdot t^2 -  2 \pi \ChirpFreqStart \cdot t + \\
        & \hspace{.1cm} 2\pi \ChirpFreqStart \cdot \tDelay - \pi \ChirpSlope (t^2 - 2t\cdot \tDelay + \tDelay^2) ] \} + z'(t)\\
        = & A_{\textrm{Rx}} \cdot \exp \{j[2\pi \ChirpSlope \cdot \tDelay \cdot t + 2\pi \ChirpFreqStart \cdot \tDelay - \\
        & \hspace{.1cm} \pi \ChirpSlope \cdot \tDelay^{2}] \} + z'(t),
    \end{split}
\end{equation}
where $z'(t) = x(t)z^{*}(t)$ represents the noise present after the mixing. From \eqref{eq_IF_signal_unsimplified}, several simplifications can be made. First, $2\pi \ChirpSlope \cdot \tDelay \cdot t$ can be simplified by recognizing that $R(t) << \RangeTarget$ and defining  $\FreqIF$ as
        \begin{equation}\label{eq_f_IF}
                \FreqIF := \frac{2 \ChirpSlope \cdot \RangeTarget}{\LightSpeed}.
        \end{equation}
        The simplified term can then be expressed as
        \begin{equation} \label{eq_IF_signal_term1_simplified}
            \begin{split}
                2\pi \ChirpSlope \cdot \tDelay \cdot t &
                = 2 \pi \ChirpSlope \left( \frac{2(R(t) + \RangeTarget)}{\LightSpeed} \right)t, \\
                & \approx 2 \pi \frac{2 \ChirpSlope \cdot \RangeTarget}{\LightSpeed} t, \\
                & = 2 \pi f_{IF} \cdot t.
            \end{split}
        \end{equation}
Next, $2\pi f_{\LightSpeed} \cdot \tDelay$ can be simplified by sampling $R(t)$ at each chrip using $R(l \cdot \ChirpDuration) = \VelocityTarget \cdot l \cdot \ChirpDuration$ where $l$ is the chirp number in the frame. Additional we define $\DopplerShift$ as
    \begin{equation}
        \label{eq_Phi_doppler}
            \DopplerShift = \frac{4 \pi \VelocityTarget \cdot \ChirpDuration}{\lambda}.
    \end{equation}
    Using these simplifications, $2\pi \ChirpFreqStart \cdot \tDelay$ simplifies as follows:
    \begin{equation} \label{eq_IF_signal_term2_simplified}
        \begin{split}
            2 \pi \ChirpFreqStart \cdot \tDelay  = & 
            2 \pi \ChirpFreqStart \frac{2(R(l \cdot \ChirpDuration) + \RangeTarget)}{\LightSpeed} \\
            = & \frac{4 \pi (R(l \cdot \ChirpDuration) + \RangeTarget)}{\lambda} \\
            = & \frac{4 \pi R(l \cdot \ChirpDuration)}{\lambda} + \frac{4 \pi \RangeTarget}{\lambda} \\
            = & \frac{4 \pi \VelocityTarget \cdot l \cdot \ChirpDuration}{\lambda} + \frac{4 \pi \RangeTarget}{\lambda} \\
            & = \DopplerShift \cdot l + \frac{4 \pi \RangeTarget}{\lambda}.
        \end{split}
    \end{equation}
Finally, $\pi \ChirpSlope \cdot \tDelay^{2}$ can be simplified by recognizing that $R(t)^{2} << 2R(t) \RangeTarget << \RangeTarget^{2}$. The simplified term can then be expressed as 
        \begin{equation} \label{eq_IF_signal_term3_simplified}
            \begin{split}
                \pi \ChirpSlope \cdot \tDelay^{2} = &
                \pi \ChirpSlope \left( \frac{2(R(t) + \RangeTarget)}{\LightSpeed}\right)^{2} \\
                 = & \pi \ChirpSlope\frac{4}{\LightSpeed^{2}}(R(t)^{2} + 2R(t)\RangeTarget + \RangeTarget^{2} ) \\
                \approx & \frac{4 \pi \ChirpSlope \cdot \RangeTarget^{2}}{\LightSpeed^{2}}.
            \end{split}
        \end{equation}
Applying \eqref{eq_IF_signal_term1_simplified},\eqref{eq_IF_signal_term2_simplified}, and \eqref{eq_IF_signal_term3_simplified} to \eqref{eq_IF_signal_unsimplified} gives 
\begin{equation} \label{eq_IF_signal_partially_simplified}
    \begin{split}
    s_{IF}^{(l)}(t)  = & 
    A_{RX} \cdot \exp\{j[2 \pi f_{IF} \cdot t + \DopplerShift l + \\
    & \frac{4 \pi \RangeTarget}{\lambda} -
    \frac{4 \pi \ChirpSlope \cdot \RangeTarget^{2}}{\LightSpeed^{2}}\} + z'(t).
    \end{split}
\end{equation}
where $s_{IF}^{(l)}(t)$ is the IF signal corresponding to the $l$th chirp. Finally, by simplifying and defining $A_{IF} = A_{Rx} \cdot \exp \left\{j\left[ \frac{4 \pi \RangeTarget}{\lambda} - \frac{4 \pi \ChirpSlope \cdot \RangeTarget^{2}}{\LightSpeed^{2}}\right]\right\}$, \eqref{eq_IF_signal_partially_simplified} can be rewritten in terms of a range, velocity (Doppler), and noise term as
\begin{equation}
    \label{eq_IF_signal_simplified}
        \begin{split}
            s_{IF}^{(l)}(t)  = & \\
            & \hspace{-1cm} A_{IF} \cdot 
            \exp(j 2 \pi f_{IF} \cdot t) \cdot  
            \exp(j \DopplerShift l)
            + z'(t).
        \end{split}
\end{equation}
\subsubsection{Range Resolution and Maximum Range} \label{Appendix_Preliminaries_Range_Resolution_and_Maximum_Range}
The output of an FFT (in particular, the dominant signal frequencies) of the signal from~\eqref{eq_FMCW_IF_signal_simplified}   is used to evaluate the IF frequency. Thus, the resolution of the FFT also limits the range resolution and maximum range that a radar can detect a target at. By definition, an FFT can separate two tones if they have a frequency difference greater than $\frac{1}{T}$ where $T$ is the observation period. As such, two targets can be separately identified as long as their IF frequencies are greater than $\ChirpDuration$ \cite{rao_introduction_nodate}. 

Therefore, by converting from IF frequency to distance provides us with the following range resolution
\begin{equation}
    \label{eq_range_resolution}
    \begin{split}
        \Delta f_{IF} = &
        \frac{2 \Delta d \cdot \ChirpSlope}{\LightSpeed} =
        \frac{2 \Delta d \cdot \ChirpBW}{\LightSpeed \cdot \ChirpDuration}\geq \frac{1}{\ChirpDuration} \\
        \Rightarrow & \Delta d \geq \frac{\LightSpeed}{2 \cdot \ChirpBW} 
        \Rightarrow \RangeRes = \frac{\LightSpeed}{2 \cdot \ChirpBW}.
    \end{split}
\end{equation}

On the other hand, the maximum range that a radar could detect is limited by the IF frequency bandwidth, which is also based on the ADC sampling rate used to record the IF signal~\cite{rao_introduction_nodate}. Hence, the maximum range that a radar can detect an object at is obtained as
\begin{equation} \label{eq_max_range}
    \begin{split}
        \FreqSamp \geq & f_{\textrm{IFMax}} = \frac{2 \ChirpSlope \cdot \RangeMax}{\LightSpeed} 
        \Rightarrow
        \RangeMax = \frac{\FreqSamp \cdot \LightSpeed}{\ChirpSlope}.
    \end{split}
\end{equation}

\subsubsection{Velocity Resolution and Maximum Velocity} 
\label{Appendix_Preliminaries_Velocity_Resolution_and_Maximum_Velocity}
First, we observe that the phase shift caused by an object's velocity must not be greater than $\pi$ between successive chirps so as not to be ambiguous. Thus, the maximum velocity satisfies
satisfies~\cite{rao_introduction_nodate}
\begin{equation} \label{eq_max_velocity}
    \DopplerShift \leq \pi \hspace{10pt} \Rightarrow \hspace{10pt} 
    v \leq \frac{\lambda}{4 \ChirpDuration}.
\end{equation}

%Next, we can use a 
A similar approach can be used to determine the velocity resolution. In radians, two tones are separable if $\Delta \DopplerShift \geq \frac{2\pi}{\NumChirpsPerFrame}$, where $\NumChirpsPerFrame$ is the number of chirps. %Using this, we can derive an equation \ref{eq_A1_velocity_resolution} for 
Hence, the velocity resolution satisfies~\cite{rao_introduction_nodate}
\begin{equation} \label{eq_velocity_resolution}
    \begin{split}
        \DopplerShift &= \frac{4 \pi \VelocityTarget \cdot \ChirpDuration}{\lambda}, \\
        \Delta \DopplerShift &\geq \frac{2 \pi}{\NumChirpsPerFrame} \hspace{10pt}
        \Rightarrow 
        \hspace{10pt}
        \Delta v \geq \frac{\lambda}{2 \NumChirpsPerFrame \ChirpDuration},
    \end{split}
\end{equation}
where $\NumChirpsPerFrame$ is the number of chirps in a radar frame.

%% file: A2_Appendix_2.tex
\subsection{Simulation Details} \label{Appendix_Simulation_Details}

\subsubsection{Distribution of Test Cases for Simulated Victim Radar Parameter Estimation}
\label{Appendix_Simulation_Details_Sensing_Component_Testing}

\begin{figure}[!t]
    \centering
    \includegraphics[width=0.994\columnwidth]{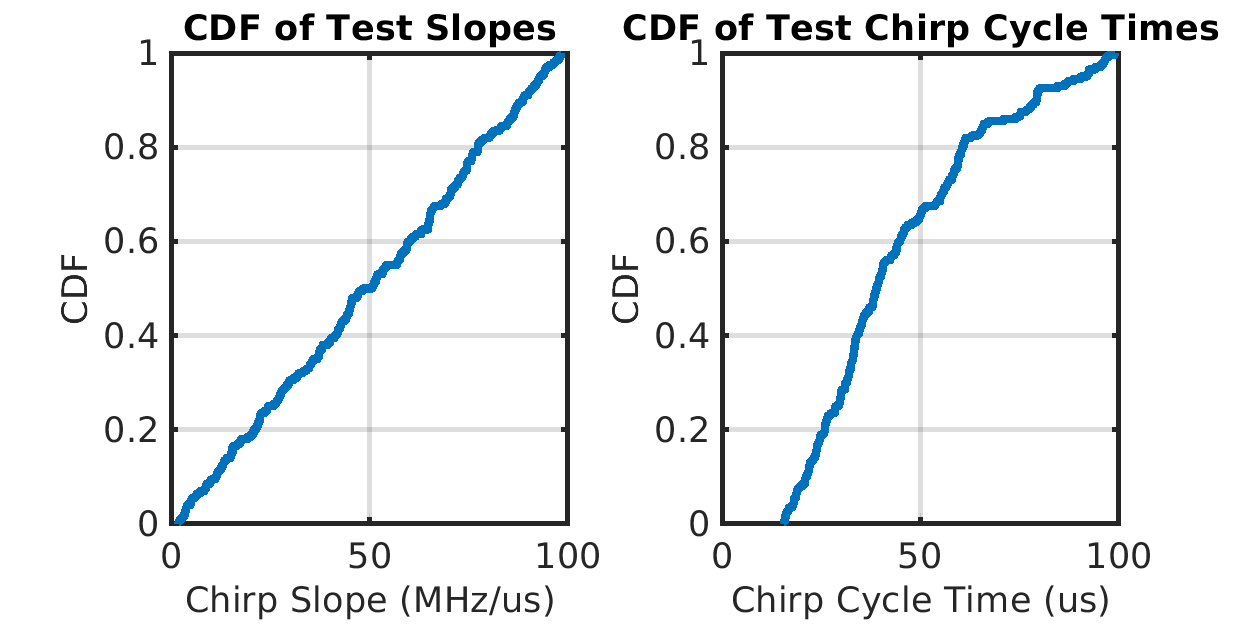}
    \caption{Test case distribution for the simulation-based evaluation of the victim radar parameter estimation.}
    \label{fig_simulation_sensing_evaluation_test_cases_cdf}
\end{figure}

Fig.~\ref{fig_simulation_sensing_evaluation_test_cases_cdf} shows the distribution of chirp periods and chirp slopes used for the simulation-based evaluation of the victim parameter sensing presented in Section~\ref{Victim_Parameter_Sensing}.

\subsubsection{Distribution of Test Cases for Attacker Spoofing Accuracy}
\label{Appendix_Simulation_Details_Spoofing_Tests}

\begin{figure}
    \centering
    \includegraphics[width=0.994\columnwidth]{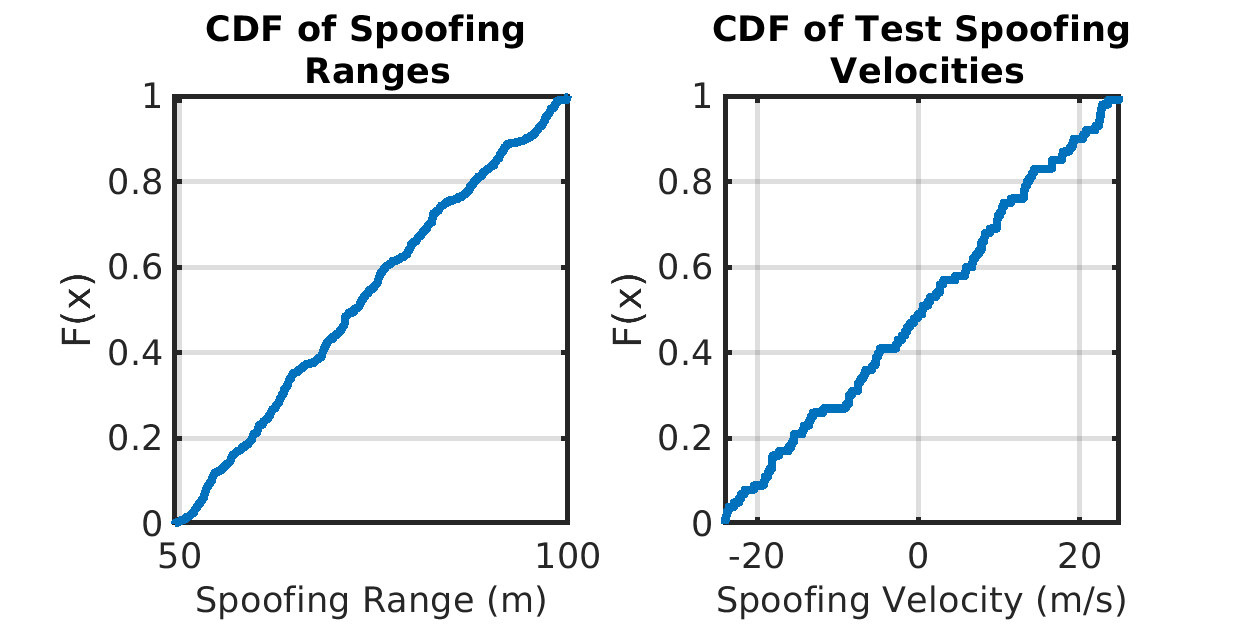}
    \caption{Test case distribution for the simulation-based spoofing accuracy evaluation.}
    \label{fig_simulation_spoofing_test_cases_cdf}
\end{figure}

The distributions of the desired spoofing ranges ($\RangeSpoof$) and velocities ($\VelocitySpoof$) used for the simulation-based evaluation of the attack spoofing accuracy in Section~\ref{Large_Scale_Evaluation} are presented in Fig.~\ref{fig_simulation_spoofing_test_cases_cdf}.

\subsubsection{CFAR Detection Regions for Large-Scale Evaluation}
\label{Appendix_Simulation_CFAR_Detection_Region}

The CFAR detection regions for each victim radar configuration utilized in the simulation-based large scale evaluations from Section~\ref{Large_Scale_Evaluation}, are listed in Table~\ref{table_simulated_victim_cfar_detection_regions}.

\begin{table}[!t]
    \caption{CFAR detection regions for considered victim configurations}
    \label{table_simulated_victim_cfar_detection_regions}
    % \vspace{-6pt}
    \centering
    \begin{tabular}{ccc}
    \toprule
    Configuration & Range Region (m) & Velocity Region (m/s)\\
    \midrule
    A & 44.97 to 427.18 & -33.94 to 34.55 \\
    B & 19.26 to 252.53 & -35.23 to 35.54 \\
    C & 2.14 to 216.91 & -39.73 to 40.075 \\
    D & 0.54 to 222.62 & --39.72 to 40.073 \\
    \bottomrule
    \end{tabular}
\end{table}

%% file: A3_Appendix_3.tex
\subsection{Details for Test Cases Used for Physical Evaluation}
\label{Appendix_Experimental_Details}

\subsubsection{Distribution of Test Cases for Physical Evaluation of the \name~Sensing Accuracy}

\begin{figure}[!t]
    \centering
    \includegraphics[width=0.994\columnwidth]{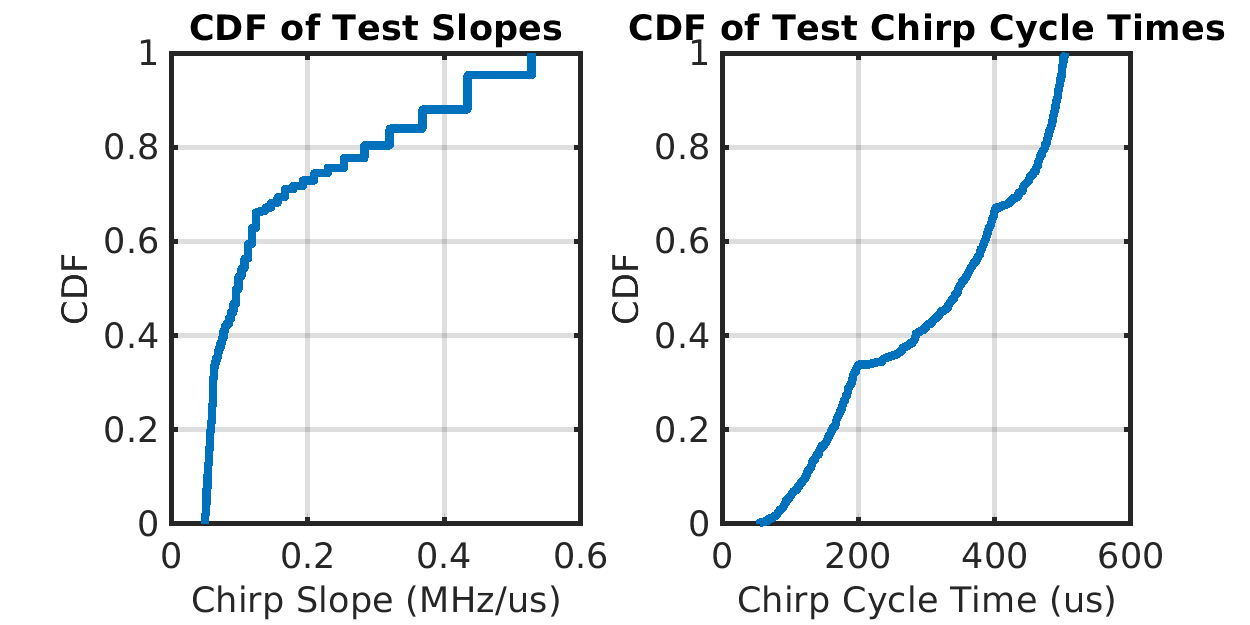}
    \caption{Distribution of test case used for experimental evaluation of the accuracy of victim radar parameter estimation.}
    \label{fig_experimental_test_cases_cdf}
\end{figure}

Fig.~\ref{fig_experimental_test_cases_cdf} presents the distributions of the desired victim radar chirp periods ($\ChirpDuration$) and chirp slopes ($\ChirpSlope$) used for the experimental evaluations performed in Section~\ref{Physical_Implementation}. To generate samples at various chirp cycle times and chirp slopes, we performed 200 trials for each of the following three groups: chirps with periods in the range [\usec{50}, \usec{200}], chirps with periods in the range [\usec{200}, \usec{400}], and chirps with periods in the range [\usec{400}, \usec{500}]. This was done to validate our real-time experimental sensing component using a wide range of victim configurations, and it explains the three distinct groupings that appear in the distribution.
%\todo{Check+check titles+ fix units -- should not be italic} 

\subsubsection{Distribution of test cases used for physical evaluation of the attack spoofing accuracy.}

\begin{figure}[!t]
    \centering
    \includegraphics[width=0.994\columnwidth]{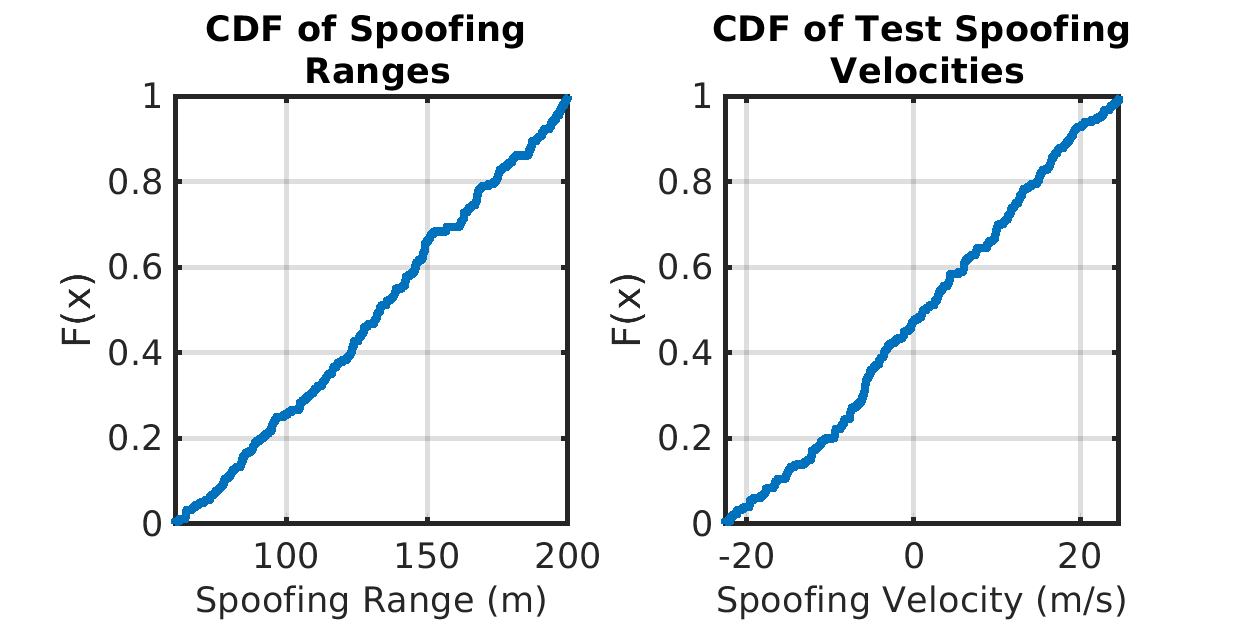}
    \caption{Distributions of test cases used for experimental evaluation of {\name} spoofing accuracy.}\label{fig_experimental_spoofing_test_cases_cdf}
\end{figure}

Fig.~\ref{fig_experimental_spoofing_test_cases_cdf} presents the distribution of the desired spoofing ranges ($\RangeSpoof$) and velocities ($\VelocitySpoof$) used to experimentally evaluate the spoofing accuracy of the {\name} prototype framework in Section~\ref{Physical_Implementation}.
%\todo{check+figure captions}